\documentclass[useAMS,usenatbib]{mn2e}
\usepackage{epsfig}
\usepackage{lscape}
\usepackage{epsfig}
\usepackage{rotating}
\usepackage{setspace}
\usepackage[maxfloats=25]{morefloats}
\usepackage[normalem]{ulem}

\title[Spectroscopic signatures of youth in low-mass kinematic candidates of young moving groups]{Spectroscopic signatures of youth in low-mass kinematic candidates of young moving groups}
\author[M. C. G\'alvez-Ortiz et al.]{M. C. G\'alvez-Ortiz,$^{1,2}$\thanks{E-mail:M.Galvez-Ortiz@herts.ac.uk} M. Kuznetsov,$^{3,4}$ J. R. A. Clarke,$^{2}$ Ya. V. Pavlenko,$^{3,2}$
\newauthor S. L. Folkes,$^{2,5}$ D. J. Pinfield,$^{2}$ H. R. A. Jones,$^{2}$  J. S. Jenkins,$^{6,2}$  J. R. Barnes,$^{2}$ 
\newauthor  B. Burningham,$^{2}$ A. C. Day-Jones,$^{6,2}$ E. L. Mart\'in,$^{1}$ A. E. Garc\'ia P\'erez,$^{7}$ 
 \newauthor  C. del Burgo$^{4}$ and R.S. Pokorny$^{8}$ \\
$^{1}$Centro de Astrobiolog\'ia (CSIC-INTA). Crta, Ajalvil km 4. E-28850, Torrej\'on de Ardoz, Madrid, Spain \\
$^{2}$ Centre for Astrophysics Research, Science and Technology Research Institute, University of Hertfordshire, Hatfield AL10 9AB, UK\\
$^{3}$ Main Astronomical Observatory, Academy of Sciences of Ukraine, Golosiiv Woods, Kyiv-127, 03680, Ukraine\\
$^{4}$ Instituto Nacional de Astrof\'\i sica, \'Optica y Electr\'onica (INAOE), Luis Enrique Erro 1, Sta. Ma. Tonantzintla, Puebla, Mexico\\
$^{5}$ Departamento de F\'isica y Astronom\'ia, Facultad de ciencias, Universidad de Valpara\'iso, Ave. Gran Breta\~na 1111, Playa Ancha, \\ Casilla 53, Valpara\'iso, Chile\\
$^{6}$ Department of Astronomy, Universidad de Chile, Casilla Postal 36D, Santiago, Chile\\
$^{7}$ Department of Astronomy, University of Virginia, P.O. Box 400325, Charlottesville, VA 22904-4325, USA\\
$^{8}$ Yunnan Observatory, P.O. Box 110. CAS. 650011 Kunming. P. R. China.
}
\begin{document}

\date{Accepted. Received ; }

\pagerange{\pageref{firstpage}--\pageref{lastpage}} \pubyear{2011}

\maketitle

\label{firstpage}

\begin{abstract}

 We present a study of age-related spectral signatures observed in 25 young low-mass 
 objects that we have previously determined as possible kinematic members of 
 five young moving groups: the Local Association (Pleiades moving group, age=20 - 150 Myr),
 the  Ursa Major group (Sirius supercluster, age=300 Myr), the Hyades
  supercluster (age=600 Myr), IC 2391 supercluster (age=35--55 Myr) and the
 Castor moving group (age=200 Myr). 
  In this paper we characterize the spectral properties of observed high or low resolution
  spectra of our kinematic members by fitting theoretical spectral distributions.
We study signatures of youth, such as lithium~{\sc i} 6708 \AA,
  H$\alpha$ emission and other age-sensitive spectroscopic signatures 
 in order to confirm the kinematic memberships through age constraints.
   We find that 21 (84\%) targets show spectroscopic signatures of youth in agreement with
 the age ranges of the moving group to which membership is implied. For two further 
 objects, age-related constraints remain difficult to determine from our analysis.
  In addition, we confirm two moving group kinematic candidates as brown dwarfs.

\end{abstract}


\section{Introduction}

Identifying members of known moving groups (MG) or open clusters
 can provide an important constraint on their age and composition.
 Low-mass stars and brown dwarfs (BDs) members of known MGs or open clusters 
 can provide important feedback to atmospheric and evolutionary models of such objects
  at young ages, which currently await good calibration from constraints on age and 
 composition. Therefore, there exists an imperative to compile a sample of MG members 
 in this mass range that can be used as anchor points or test-beds for these models.

 In this paper we present the last in a series of results of a major survey of Ultra-Cool Dwarfs
 (UCDs; objects with a spectral classification of M7 or later corresponding to Teff$<$2500 K),
 in young moving groups which began with the results presented 
 in Clarke et al. (2010), hereafter Paper I,
 and G\'alvez-Ortiz et al. (2010), hereafter Paper II.
 We focus here on the study of age-sensitive
 spectroscopic signatures such as gravity sensitive
  features (e.g. Gorlova et al. 2003; McGovern et al. 2004),
 H$\alpha$ emission (e.g. West et al. 2004, 2008, 2011) and 
 the presence of Li~{\sc i} 6708 \AA \ in a sample of objects previously
  classified as members of a MG based on kinematic or astrometric
  criteria.
 
 Despite the recent disagreement about the origin of moving groups
 (Famaey et al. 2007, 2008; Antoja et al. 2008; Klement et al.
 2008; Francis \& Anderson 2009; Zhao et al. 2009; L\'opez-Santiago et al. 2009; Bovy \& Hogg 2010; 
  Murgas et al. 2013), and as we discussed in Paper II,
 we are assuming the classical concept: a moving group is a young stellar 
 population that shares a common space motion (e.g. Pinfield et al. 2006)
 whose members have a common origin, and therefore, age and composition.
 We focused our studies on well-documented groups: the Hyades supercluster (HY; 600 Myr), 
 the Ursa Major group (Sirius supercluster) (Si; 300 Myr), IC 2391 supercluster (IC; 35-55 Myr),  
 the Castor Moving Group (CA; 200 Myr), and the Local Association
 (20-150 Myr) or Pleiades (PL) moving group (See Paper I and II for details of
 MG properties and references thereof).

 Many efforts have been made to obtain accurate models to understand the cool and complex 
 atmospheres of low mass stellar and substellar objects (e.g. Kirkpatrick et al. 1993; 
 Rajpurohit et al. 2011; Reyle et al. 2011; Rajpurohit et al. 2012). But the models still
 show discrepancies, for example in the strength of some absorption bands: discrepancies likely 
 due to inaccurate atomic parameters and/or missing molecular opacities.
 Also, dust formation and its behaviour with temperature changes spectral characteristics in many ways. 
 Models that describe Very Low-Mass stars (VLMS) and BD atmospheres cannot hope to be
 in agreement with observations without the inclusion of dust.
 The onset of iron and silicates dust grain formation suspended in the photospheres 
 of late-type M dwarfs through the L dwarf spectral sub-types is accompanied by spectral 
 reddening, especially in near-IR (based upon the size, quantity and distribution of this dust). 
 The T dwarf spectral sub-type sequence demonstrates a reversal to bluer colours due to the dust
  settling below the photosphere, thus reducing the reddening. Consequently, atmospheric models 
 that describe dwarf objects with T$_{\rm eff}$ $\leqslant$ 2800 K should include treatment 
 for the effects of atmospheric dust evolution.

 Up to date, there are several families of stellar model atmospheres of very low mass stars. 
 These models are still incomplete or approximate in some physical properties (opacities, oscillator
 strengths for some lines and molecular bands, etc), showing few or considerable discrepancies 
 depending on the atmospheric regions, temperatures, etc, and there are also non covered areas or "gaps". 
 A comprehensive study of models tested by observations is needed to fully understand this cool atmospheres.
  Since BDs are objects occupying an intermediate position between stars and giant planets,
 studying their atmospheres can help us to better understand the processes in giant exo-planet atmospheres.
 The determination of the physical properties of VLMS and BDs is also 
 important for understanding a broad range of topics including stellar and planetary
 formation, circumstellar disks, dust formation in cool atmospheres, and the initial mass function. 

 By comparing high and low resolution optical spectra of our 25 low-mass stars and BD
 kinematic MG candidates with a grid of atmospheric models, we can obtain effective
 temperature and surface gravity estimates. 

  In Section~2 we describe the sample characteristics and selection criteria, while
 in Section~3 we present details of our observations and data reduction techniques.
 In Section~4 we explain the model atmosphere fitting process, and 
 in Section~5 we describe the age indicators and the analysis used to assess the sample 
 memberships. Finally, in Section~6 we present a brief discussion and summary of our results.

\section{Sample}

 We present here the study of two different samples.

\subsection{Sample A:}

 We used photometric and astrometric criteria to combine an extended version of 
 the Liverpool-Edinburgh High Proper Motion survey (ELEHPM; Pokorny, Jones \& Hambly 2003; Pokorny et al. 2004) 
 and Southern Infrared Proper Motion Survey (SIPS; Deacon et al. 2005; Deacon \& Hambly 2007) 
 with the Two Micron All Sky Survey (2MASS), and applied colour cuts of
 $J$-$K$$_s\geq$1.0 and $R$-$K$$_s\geq$5.0  to select objects with spectral type 
 predominantly later than M6V. To these we applied proper motion restriction, 
 $\mu/\Delta\sigma$$_{\rm{\mu}}>4$ ($\mu$ is proper motion and $\sigma_{\rm{\mu}}$ is the 
 proper motion one sigma uncertainty) to ensure a 4-$\sigma$ level of proper motion 
 accuracy before searching for stars with proper motions consistent with any of  
 the MGs. Of the 817 red object catalogue we obtained by 
 using both astrometric and photometric criteria, we concluded that 132 of these objects 
 were possible members of one or more of the five MGs in the study (see Paper I for details).

 From this sample we then obtained high resolution spectra of 68 objects with spectral types ranging M4.5-L1. 
 Using these high resolution spectra we derived galactic space-velocity components and, applied a simple kinematic 
 criterion.  We found that 49 targets have kinematics consistent with the young 
 disk population (YD; see e.g. Eggen 1984a,b, 1989) and that 36 of them possibly belong to one of 
 the five moving groups in this study. We also measured (where possible) projected rotational velocities and 
 used the rotation rate as a supporting criterion of youth in confirming the kinematic members. 
 We found that 31 young disk candidates have moderately high rotational velocities which also 
 suggest youth and that 12 others have significantly high rotations to consider for further investigation as 
 possible young stars (see Paper II for details).

 We present here high resolution follow-up observations for the 13 brightest stars in this sample,
 that we took in order to use other spectroscopic age 
 criteria to confirm kinematic membership.
 In Table~\ref{tab:samplea} we give the name, coordinates, 2MASS $J$ magnitude and the MG candidature
 assigned from the kinematic study and the rotational velocities obtained in Paper II. 

\subsection{Sample B:}

  Our second sample consists of objects selected from the catalogue of Folkes et al. (2012), 
 and the reader is refered to this publication for full details of the selection method used in the 
 creation of the catalogue.
  These authors present an UCD catalogue containing 245 objects with spectral types 
 of late-M through to the L--T transition from low southern Galactic latitudes, 
 in an area covering 5042 sq.degs within 220$^\circ$ $\leqslant$ l $\leqslant$ 260$^\circ$ 
 and 0$^\circ$ $\leqslant$ l $\leqslant$ 60$^\circ$ for $\vert$b$\vert$$\leqslant$15$^\circ$), 
 identified as part of a deep search using 2MASS near-IR and SuperCOSMOS optical photometry
  selection criteria as well as reduced proper motion constraints.

 \subsubsection{UCD sample selection procedure:}
 \begin{itemize}
 
\item Photometry: Using an optimal set of optical/near-IR selection criteria 
 for the ($J$-$H$)/($J$-$K_{s}$) two-colour plane along with optical/near-IR colours
  of ($B_{J}$-$K_{s}$), ($R_{c}$-$K_{s}$), ($I_{c}$-$K_{s}$) and the optical colour of 
 $($R$-$I$)_{c}$, with $J$$H$$K_{s}$ from 2MASS, we obtained a sample of objects in the
  spectral range of $\sim$M8V to $\sim$L9.
  The selection method also included an $R_{F}$ and $I_{N}$ photometric surface gravity 
 test with additional photometric constraints. See Figures~1, 2 and 3, and 
 Section 2 of Folkes et al. (2012) for details.

\item  Astrometry: To eliminate bright distant contaminants, which one would expect to display small 
 proper motions, a reduced proper-motion diagram was used to segregate giant from
 dwarfs (e.g. Luyten 1978). In this case we used a diagram of reduced proper motion in 
 2MASS $K_{s}$-band plotted against ($V$-$K_{s}$) colour.
 See Fig.~6 and Section 2.5 of Folkes et al. (2012) for details.

\end{itemize}

 \subsubsection{MG Selection procedure:}

 To find possible MG members from this UCD sample, Folkes (2009) applied the 
 astrometric and photometric test described in Sections 3.2 and 3.3 of Paper I (used
 in the selection process of sample A.)
 Spectral Types and photometric distances were derived from the ($I$-$J$) colour using the 
 $M_{J}$/($I_{C}$-$J$), and $M_{J}$/SpT relations of Dahn et al. (2002). 
 
 From these photometric and astrometric criteria Folkes (2009) identified 23 objects,
 with spectral types ranging from M3 to L0 as possible moving group members, from 
 which we could observe 12 that form sample B.

 Here we obtain spectral classification and confirm (or otherwise) the youth of 
 these candidates, through low resolution spectroscopy and by studying age-sensitive 
 spectral features.
 In Table~\ref{tab:sampleb} we give the name, coordinates, 2MASS $J$ magnitude and the MG candidature
 assignation. 

\subsubsection{Spectral type determination:}

 To improve the previous photometric spectral type determination from
 Folkes (2009), we calculated a variety of different spectroscopic
 indices: PC3, TiO1+TiO2 and VO1+VO2 from Mart\'in et al. (1999);
 The VO index from Kirkpatrick et al. (1995); 
 VO-a and TiO5 from Cruz \& Reid (2002). 
 Since the TiO1+TiO2 and VO1+VO2 indices can imply two different 
 spectral types (see Fig.~9 of Mart\'in et al. 1999), we only considered
 them if they were consistent with the spectral types obtained by
 the other indices. From these a mean spectral type was calculated.

 Some objects show considerable discrepancies between spectral type derived
 from photometry and that from spectroscopic indices. That is, with the exception of two objects
 (2MASS1756-4518 and 2MASS1909-1937) with very similar photometric and spectroscopic
 spectral classification. This difference is marked by the systematic 
 assignment of a later spectral type by the photometric method. Also the difference between
  the two classifications appears greater for redder objects. 
 Even with care during selection, the objects of sample B are situated in
 highly populated and reddened areas within the photometric selection planes,
  which could have produced these differences. 
 Also, the possible presence of dust in these atmospheres could account for redder colours,
  therefore suggesting later spectral types. We used the spectral types derived from 
 spectroscopy for all calculations.

\section{Observations}

 For this paper we analysed optical
 high resolution echelle spectra of 13 objects (sample A; Table~\ref{tab:samplea}) and
 low resolution long slit spectra of 12 objects (sample B; Table~\ref{tab:sampleb}). The data
 were obtained during four observing runs detailed in Table~\ref{tab:obs}.

Both sets of high and low resolution spectra were extracted using the standard reduction procedures in the
 IRAF\footnote{IRAF is distributed by the National Optical Observatory,
 which is operated by the Association of Universities for Research in
 Astronomy, Inc., under contract with the National Science Foundation.}
 {\sc twodspec} and {\sc echelle} packages respectively: bias subtraction, flat-field division,
 extraction of the spectra, telluric correction as well as wavelength and flux calibration.
  We obtained the solution for the wavelength calibration by taking spectra of a Th-Ar lamp.
 The average signal-to-noise (S/N) of the data, measured as
 the square root of the signal at $\approx$ 8100 \AA, is $\approx$30 for UVES data
 (except for 2MASS2039-1126 that has S/N of $\approx$15) and $\approx$200
 for both runs of IMACS data.

\begin{figure*}
   \centering
   \includegraphics[width=8.5cm,angle=0,clip]{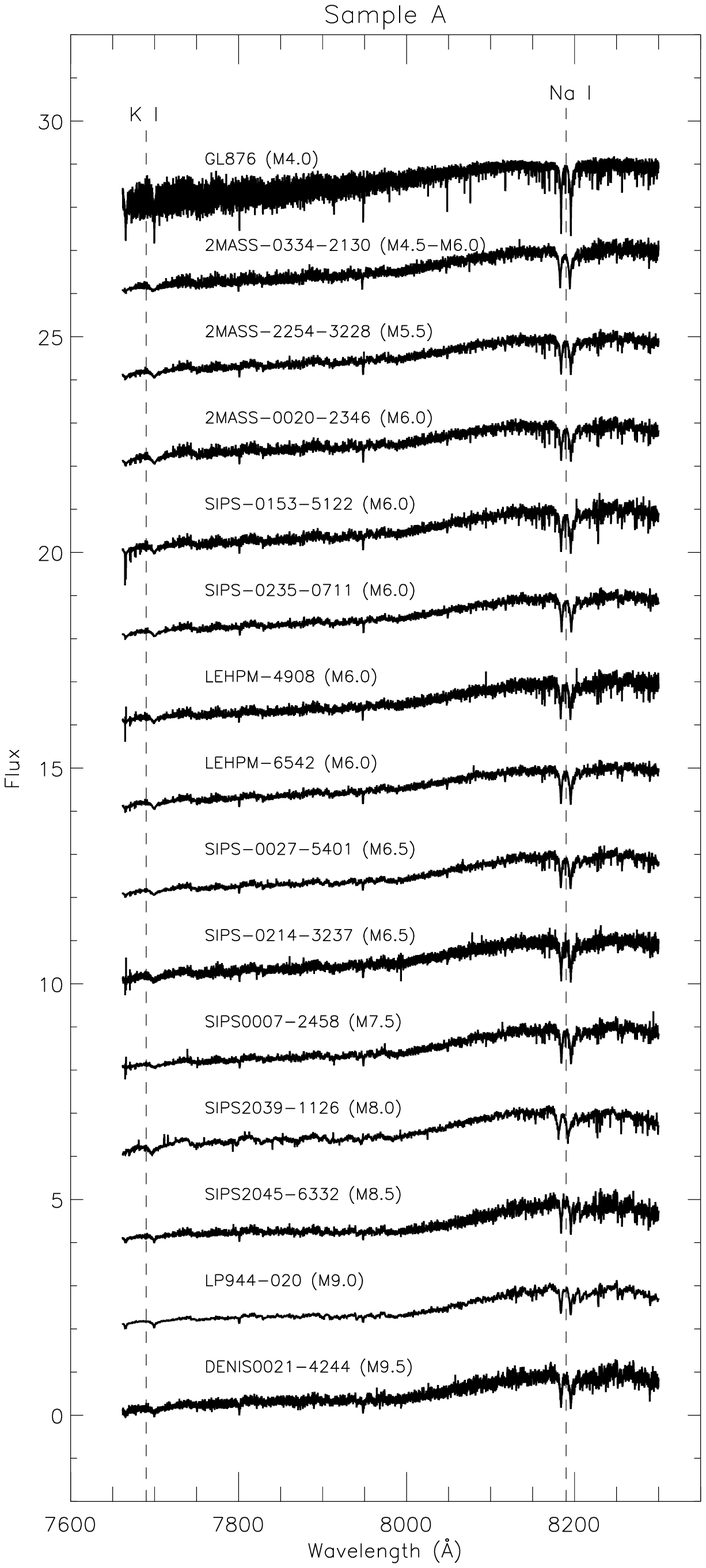}
   \includegraphics[width=8.5cm,angle=0,clip]{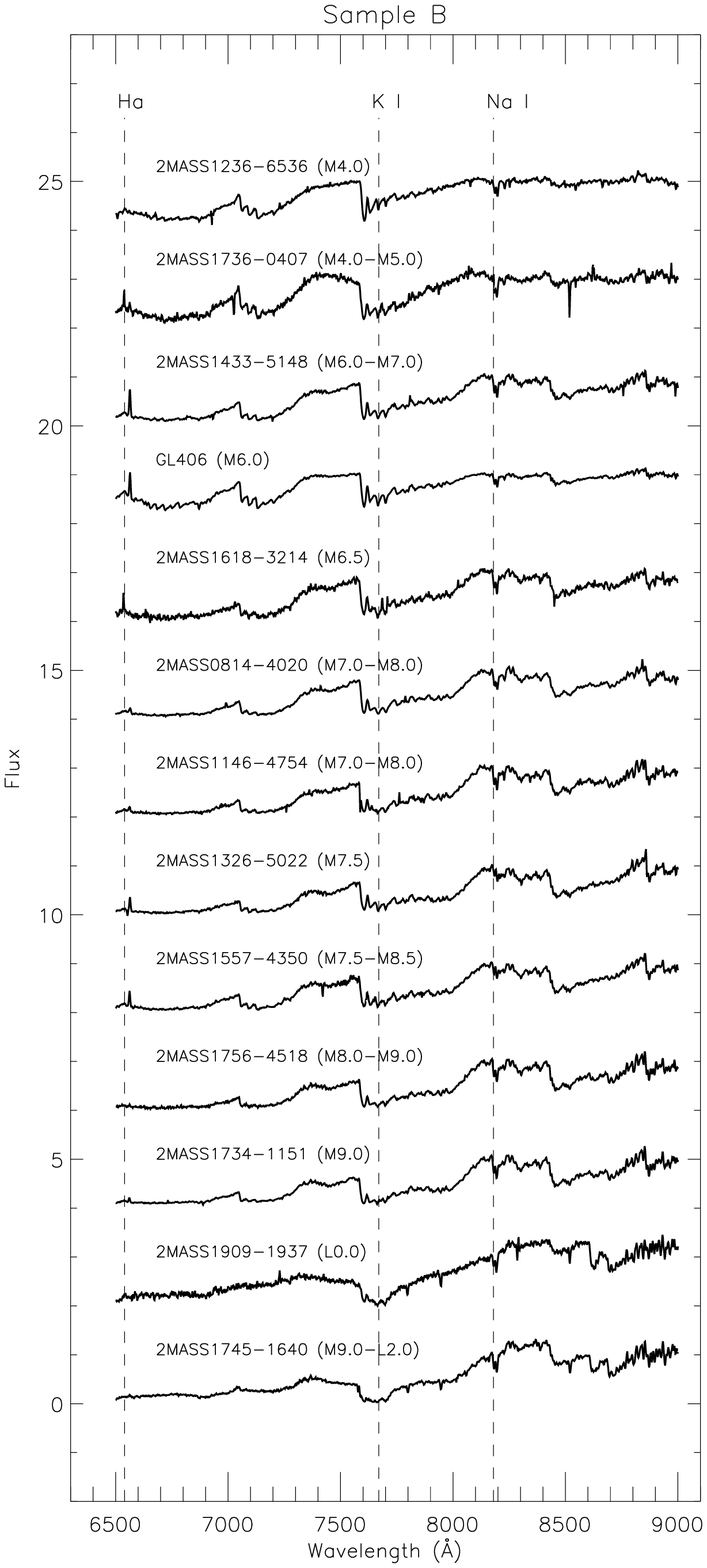}
      \caption{Left: Observed SEDs of objects from sample A, the 13 targets plus 
 reference objects in the 7600 to 8400 \AA \ range.
 Right: Observed SEDs of objects from sample B, the 12 targets plus one reference 
 object in the 6500 to 9000 \AA \ range. They are
 ordered by spectral type derived from spectroscopy.}
         \label{fig:todashl}
   \end{figure*}

 Figure~\ref{fig:todashl} shows high resolution spectra of sample A on the left and low resolution
 spectra of sample B on the right, ordered by spectral type classification.
  Well studied late-type objects with known spectral type were observed at the same time as
  the targets to be used as reference (M9 LP944-20; M4 GL876; and M6 GL 406) 
 and are also included in the figure for comparison.

\section{Atmospheric models and synthetic spectra}

 To determine spectral characteristics of the targets (T$_{\rm eff}$, $log~g$, etc),
 to aid us in age determination, 
 we compared a grid of generated synthetic spectra with the observations.  

 According to current concepts, dust plays an important role in the formation of
 late M dwarf spectra (Tsuji et al. 1996; Hauschildt \& Allard 1992). 
 Although there is no unified opinion about temperature range where dust should be considered,
 the temperatures of transition between dust-free and dusty models range from $T_{\rm eff}$$<$2600 
 (Allard et al. 2011) to $T_{\rm eff}$$<$3000 K (Jones \& Tsuji 1997).

 Therefore, synthetic spectra were computed for
 dust-free NextGen atmospheric models and for semi-empirical models
 (i.e. modified NextGen models) in the wavelength range 6500-9000 \AA .
 The ``dust effects" are included in semi-empirical models for M3-M8 stars.
 Synthetic spectra based on DUSTY, COND and corresponding semi-empirical
 models were calculated for M7-L0 stars.

\subsection{Synthetic spectra based on NextGen models}

Synthetic spectra were computed for NextGen atmospheric models 
 (Hauschildt \& Allard 1992) by the WITA6 program (Pavlenko et al. 1997; 2007a).
 The calculations were made under the assumption of local thermodynamic
 equilibrium (LTE), hydro-static equilibrium, in the absence of sources and sinks of energy,
 for a one-dimensional model atmosphere. We used the
 atomic line list from the VALD database (Kupka et al. 1999) and the line list of
 titanium oxide (TiO) from Plez (1998) to produce the synthetic spectra. 
 Vanadium oxide (VO) band opacities were computed
  in the framework of the JOLA approximation (see Lyubchik \& Pavlenko 2000).
 Synthetic spectra were calculated for a grid of models
 having T$_{\rm eff}$=2600$-$3400 K with $100$ K step, surface gravities of
 $log~g$=4.0$-$5.5 with 0.5 step and metallicities
  of $[M/H]$=0.0, -0.5 and -1.0 dex. Solar abundances were taken from Anders \& Grevesse (1989).
  The artificial rotational-broadening of spectral lines was implemented following
 the methodology of Gray (2005) using the rotational velocity of the objects obtained in Paper II. 
 We carried out a few numerical experiments to investigate the effect of microturbulence velocity 
 on the spectra. The study shows that the differences in synthetic spectra associated with $v_{t}$ in 
 the 1.0-5.0 km s$^{-1}$ range are negligible.
 Thus all theoretical spectra were computed for a microturbulence velocity $v_{t}$ = 3.0 km s$^{-1}$.

\subsection{Semi-empirical atmospheric models}

 We used the semi-empirical atmospheric models described in Pavlenko et al. (2007a) to 
 include the effects of dust formation. The semi-empirical models were obtained
 by modifying NextGen, DUSTY and COND models. 

 The dust opacity originates in the shell-like 
 structures lying above the photosphere (Pavlenko et al. 2007a) and we
 expect the influence of dust in a spectrum to be more prevalent in the shortest
 wavelength spectral region.
 We assumed that dust clouds are located in the upper most layers 
 of stellar atmosphere and do not affect the distribution of temperatures and pressures.
 Dusty effects were treated in two ways: (1) decreasing molecular abundance in 
 the gas due to molecular condensation of dust particles and (2) radiation scattering in 
 the dust clouds. The decrease in the concentration of TiO and VO molecules was modelled by two parameters: 
 (1) the number of layers in the above model for which there is no absorption of TiO (i.e., all  
 titanium oxide condenses on dust particles) and (2) the coefficient of TiO molecular density reduction 
 for all other model layers. 
 Radiation scattering in the dust clouds was also modelled by two parameters: 1) the optical thickness 
 of the dust cloud and 2) the location of maximum opacity as a result of these clouds. 
 This method is described in Pavlenko et al. (2007a) and Kuznetsov et al. (2012).

 The theoretical spectra were computed using a semi-empirical atmospheric model for a grid 
 with effective temperatures of T$_{\rm eff}$=2600$-$3000 K (based on NexGen models) 
 and $T_{\rm eff}$=2000$-$2600 K (based on DUSTY and COND models) with a step of 100 K, surface gravities of
 $log~g$=4.0$-$5.5 with a step of 0.5, and solar (Anders \& Grevesse 1989) metallicity.
 These models were calculated for the same rotational and microturbulence velocity
  as for the NextGen models. 
 We are thus able to compare the observed high and low resolution spectra with 
 synthetic spectra assuming both the presence and the absence of dust.

\subsection{Selection of best fit parameters. Objects with M4-M8 spectral type} 

   The $S$-function analysis described in Pavlenko et al. (2007a) 
 was used to determine the best fit for each grid of the model independently, 

 $$ S=\sum(f_h * H_{{\rm synt}}-H_{{\rm obs}})^2 $$

 where $f_h$ is a normalization parameter, $H_{{\rm synt}}$ is the synthetic flux 
 and $H_{{\rm obs}}$ is the observed flux.
 The minimum value of $S$-function corresponds to the best fit.

 The $S$-function is integrated in the 6900-7200 \AA, 8100-8400 \AA \ and 7660.8-7734 \AA, 8160-8220 \AA \
 wavelength ranges for low and high resolution spectra respectively. The use of these regions allowed 
 us to analyze the TiO band at $\sim$7100 \AA \, and Na~{\sc i} (8183 \AA, 8199 \AA) doublet in low 
 resolution spectra, the  K~{\sc i} (7665 and 7669 \AA) resonance doublet and Na~{\sc i} 
 (8183 \AA, 8199 \AA) subordinate doublet in high resolution spectra. 
 Analysis of Na~{\sc i} and K~{\sc i} lines is significant because their profile is
 sensitive to T$_{\rm eff}$ and $log~g$.
 The VO ($\sim$7400 \AA, $\sim$7800 \AA) bands were excluded from the fitting procedure since we do not have 
 a satisfactory line list for this molecule.

\begin{figure*}
   \centering
   \includegraphics[width=6.0cm,angle=270,clip]{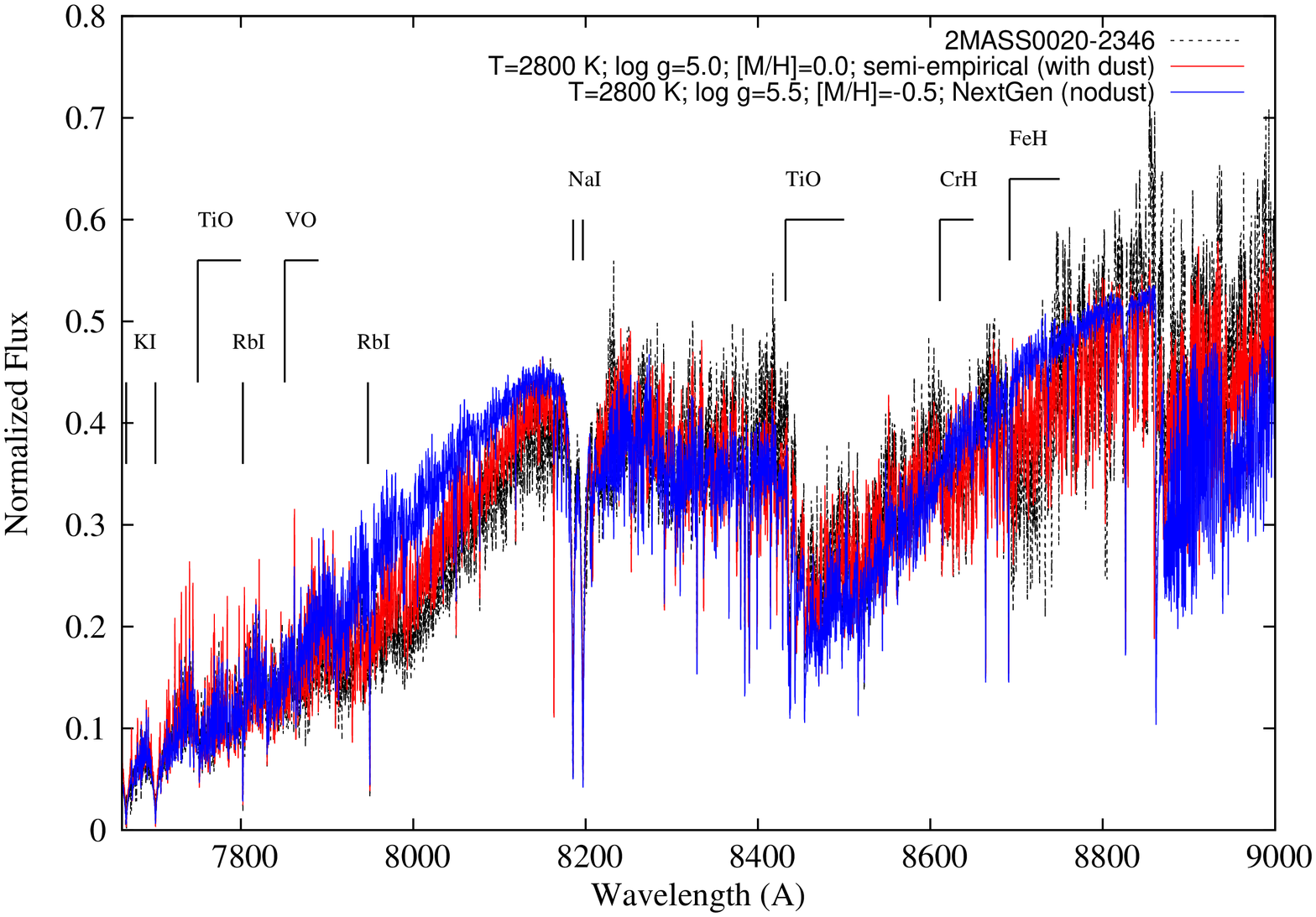}
   \includegraphics[width=6.0cm,angle=270,clip]{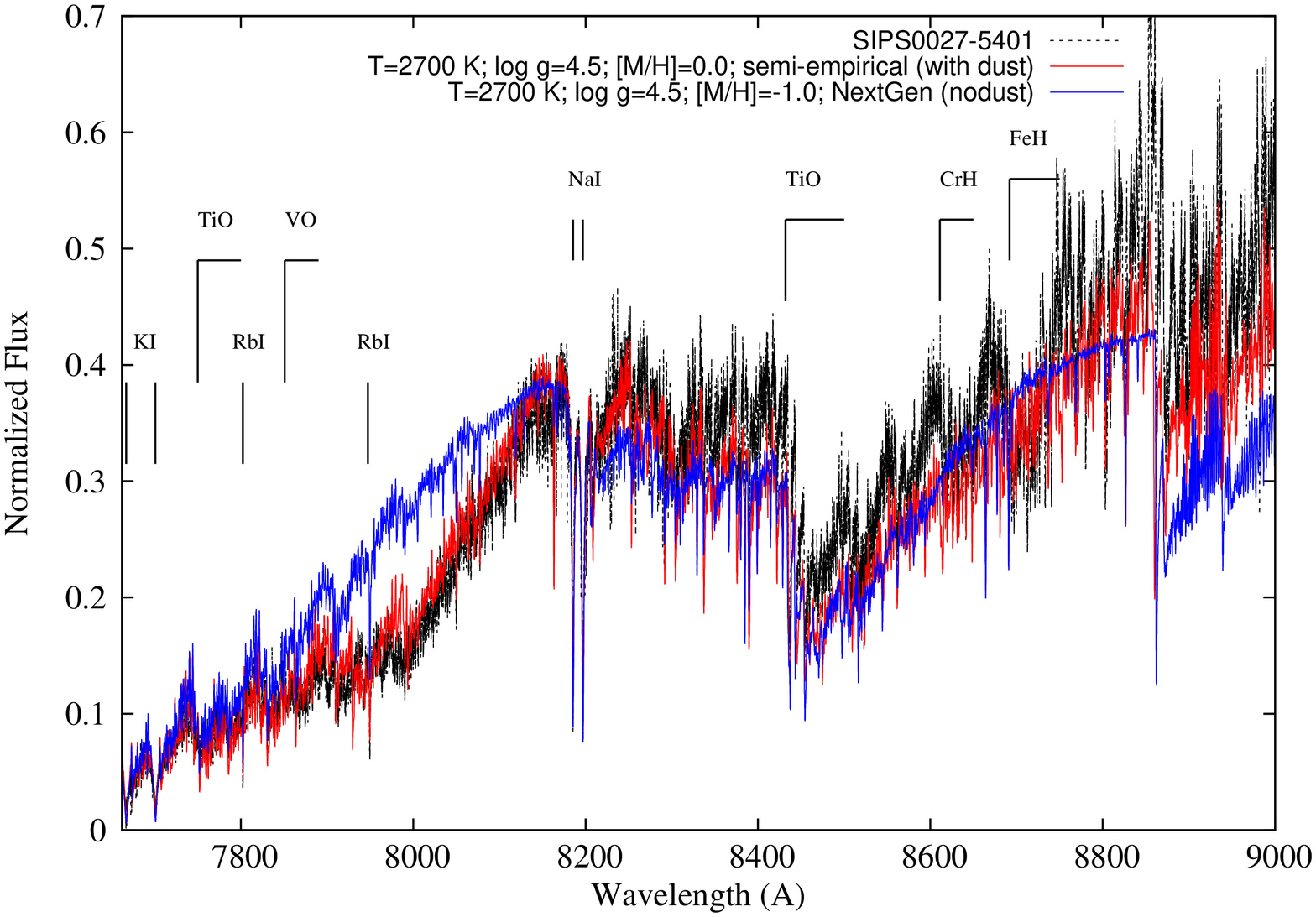}
      \caption{Sample A: random example of fits of theoretical spectra to the observed SEDs.}
         \label{fig:fight}
   \end{figure*}

\begin{figure*}
   \centering
   \includegraphics[width=6.0cm,angle=270,clip]{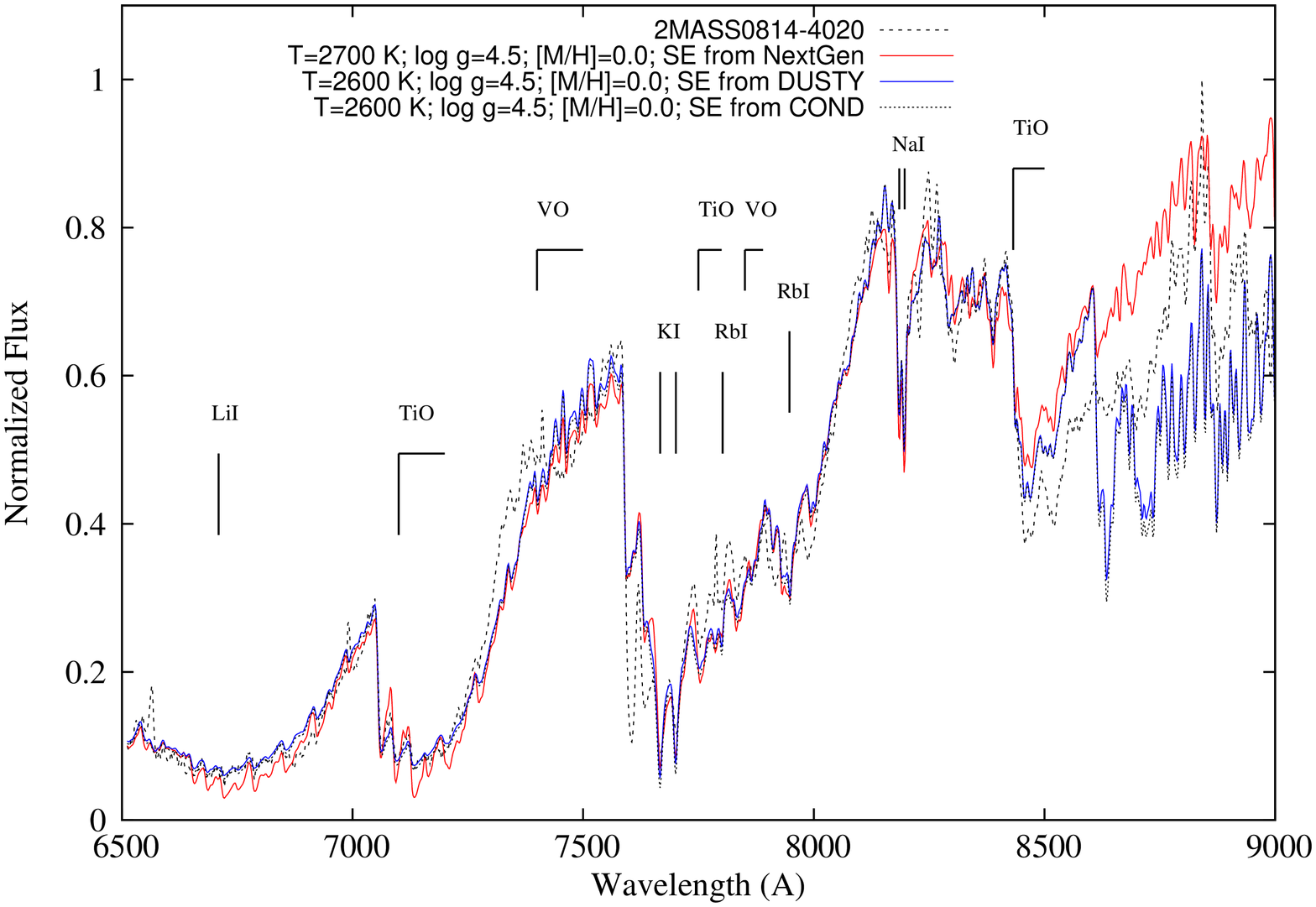}
   \includegraphics[width=6.0cm,angle=270,clip]{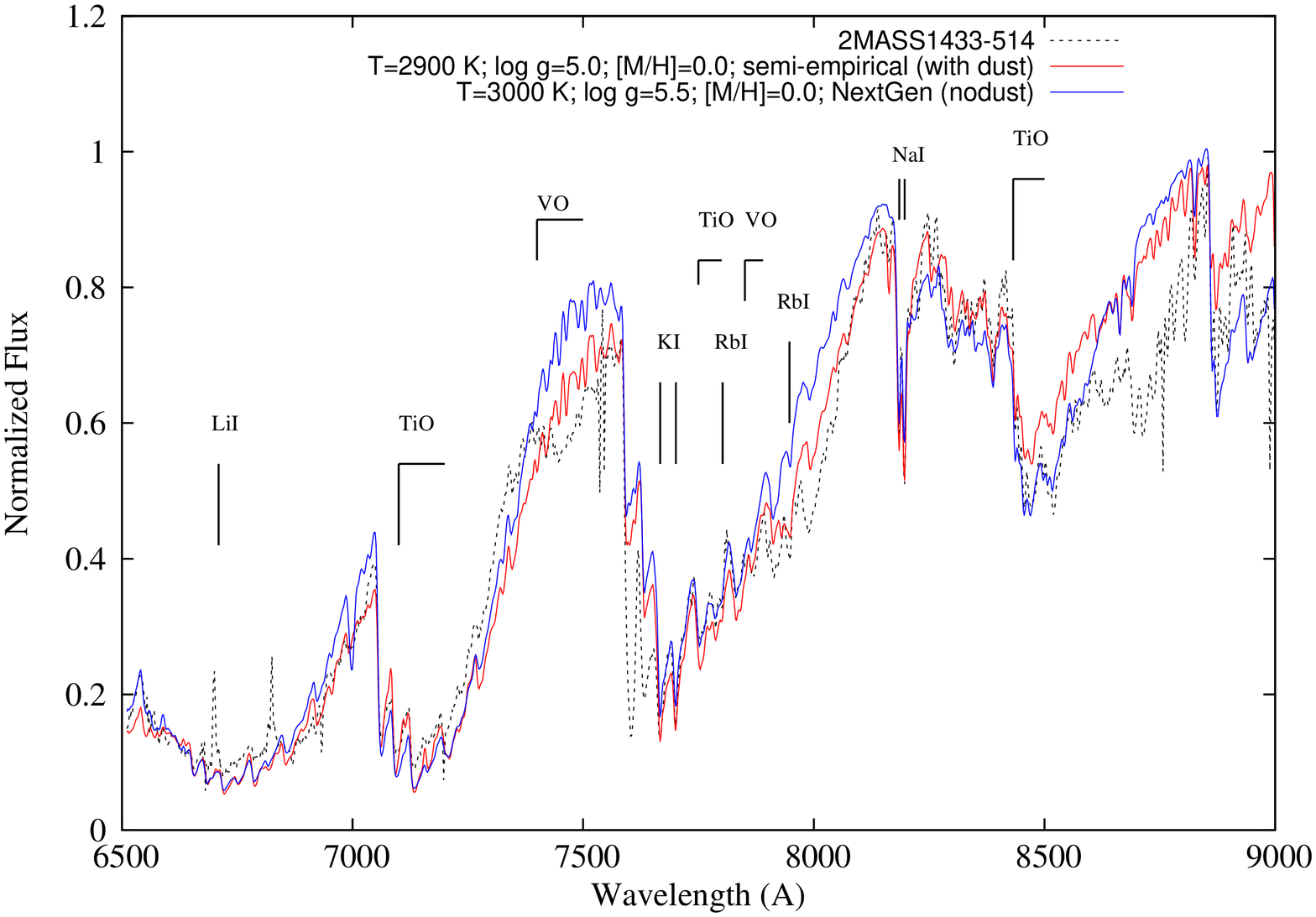}
      \caption{Sample B: random example of the best fits to the observed SEDs.
              }
         \label{fig:figlt}
   \end{figure*}

\begin{figure*}
   \centering
   \includegraphics[width=6.0cm,angle=270,clip]{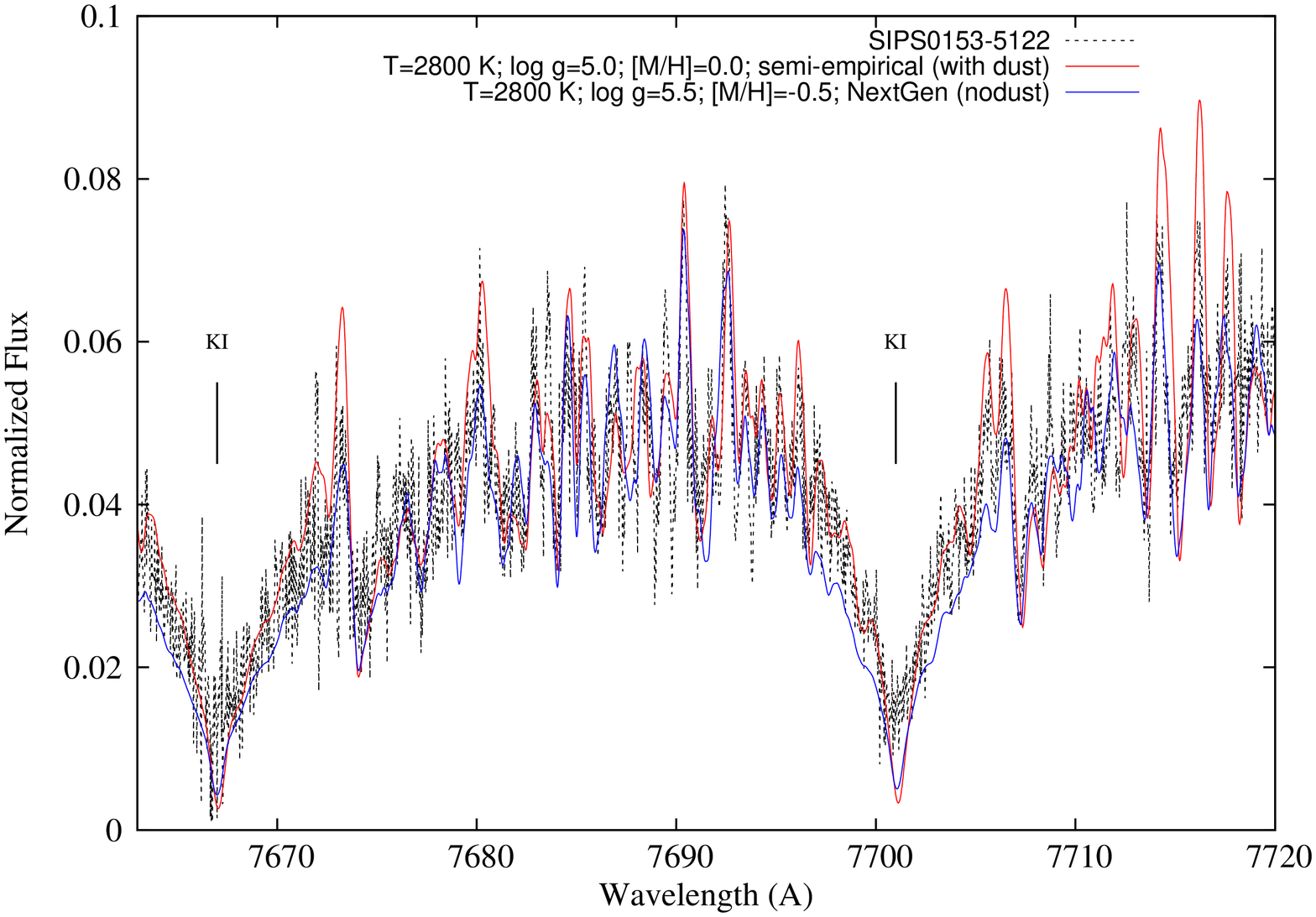}
   \includegraphics[width=6.0cm,angle=270,clip]{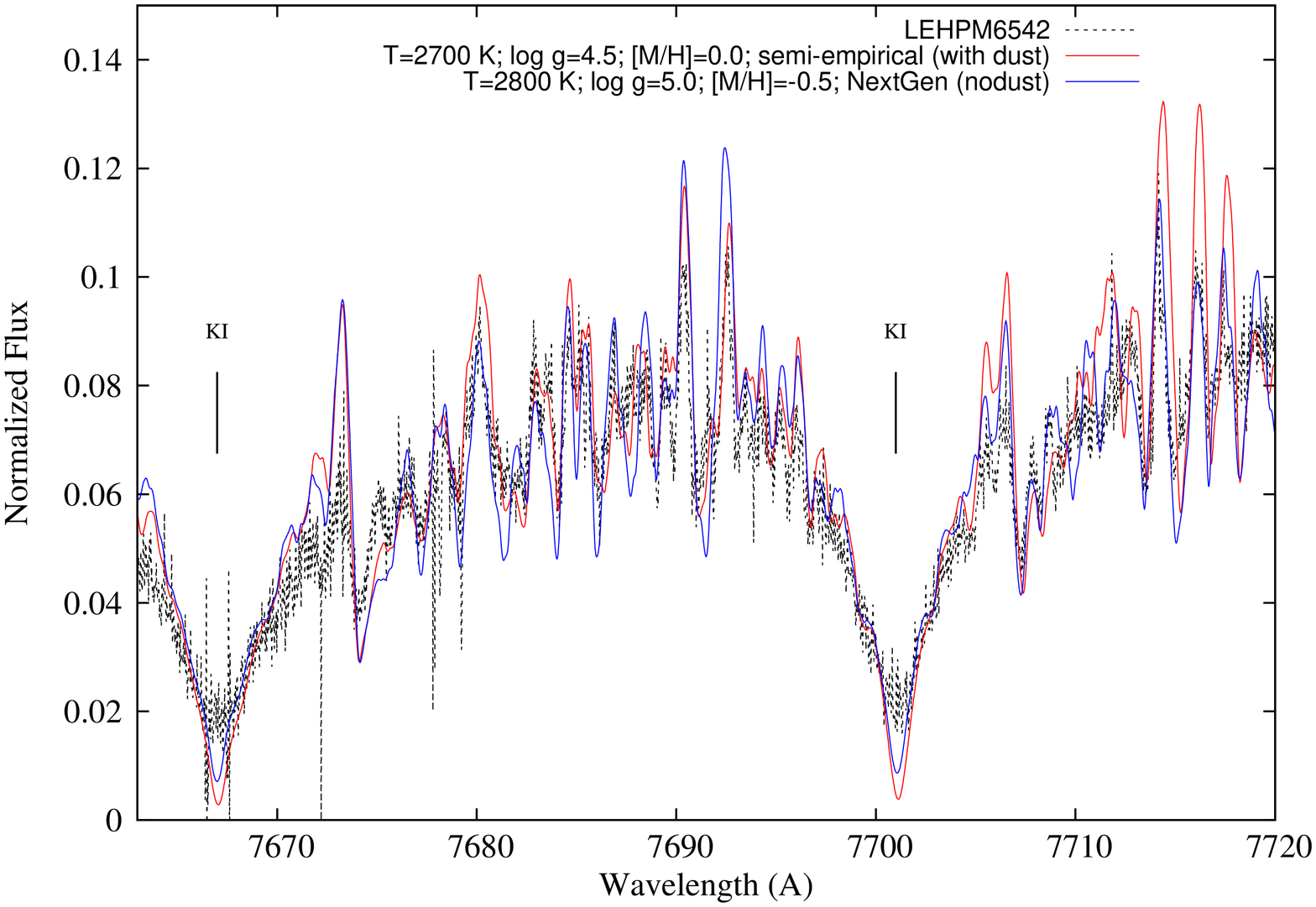}
      \caption{Sample A: random example of the best fits to the spectral regions across K~{\sc i} lines.
        }
         \label{fig:fighk}
   \end{figure*}

\begin{figure*}
   \centering
   \includegraphics[width=6.0cm,angle=270,clip]{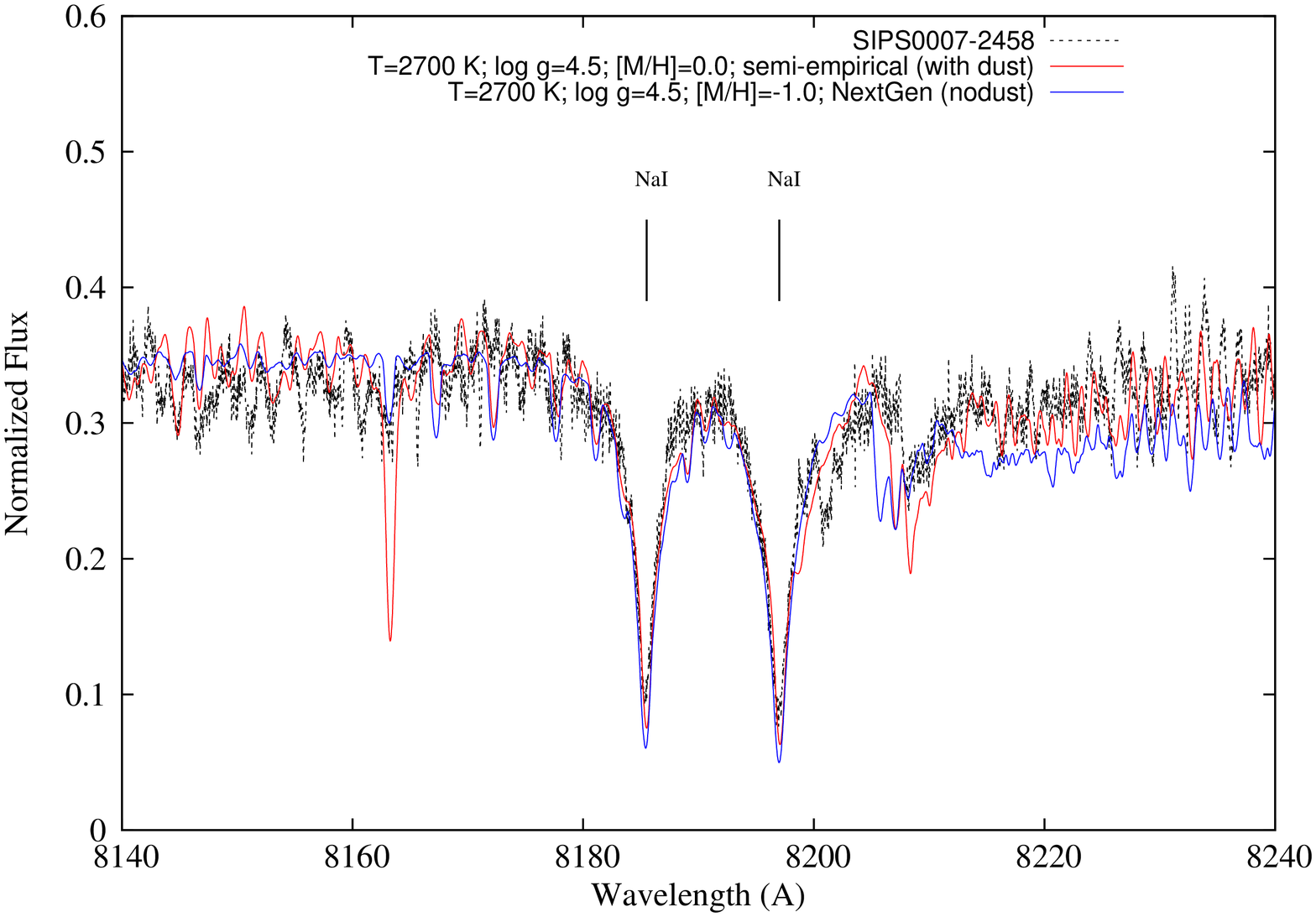}
   \includegraphics[width=6.0cm,angle=270,clip]{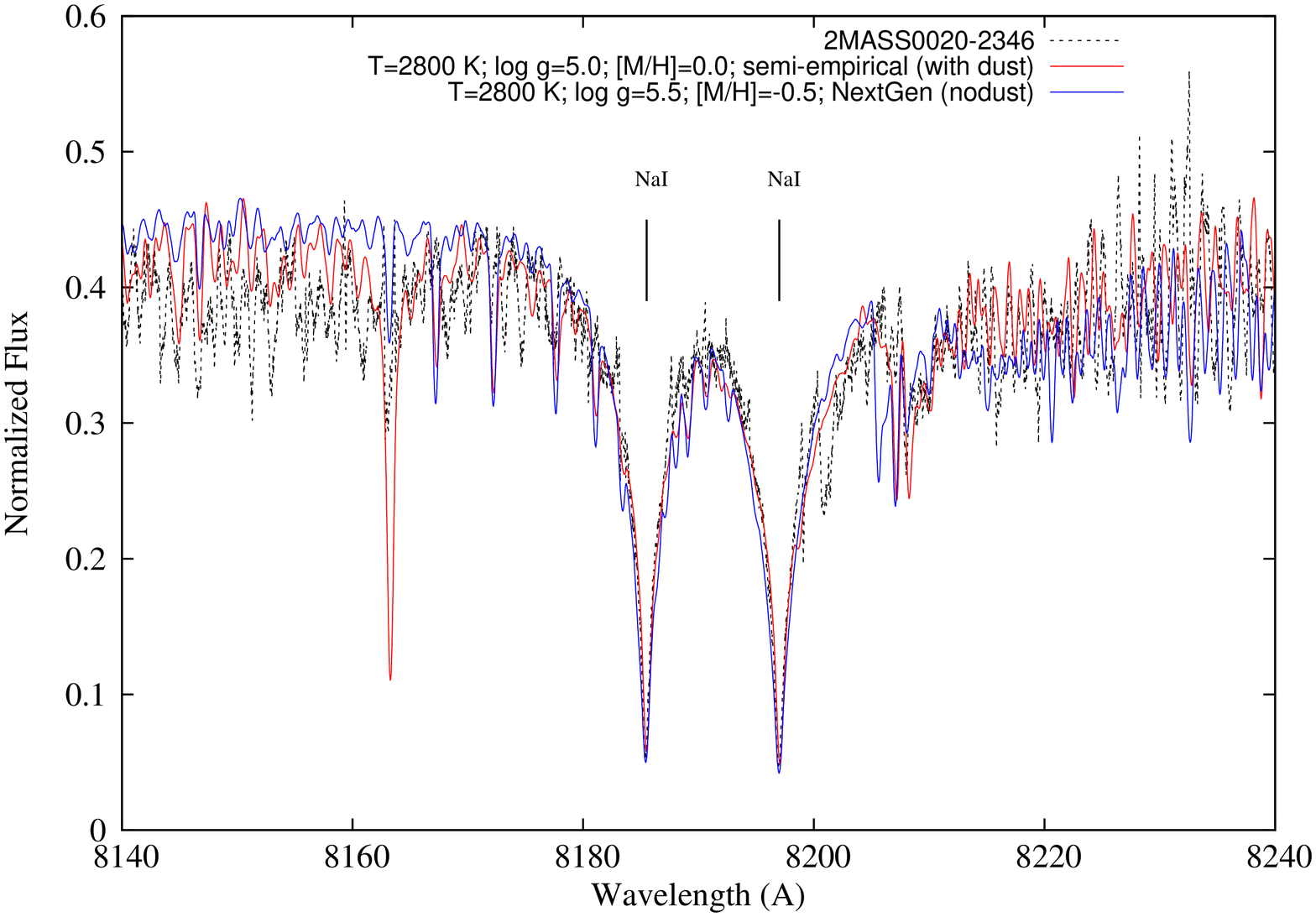}
      \caption{Sample A: random example of the best fits to the spectral regions across Na~{\sc i} lines.
              }
         \label{fig:fighna}
   \end{figure*}

 We carried out fits of the observed spectra for two model grids, original
 NextGen and semi-empirical, for M4-M8 objects.
  Figures~\ref{fig:fight} and \ref{fig:figlt} show example fits for the
 red part of the spectra in sample A, and the complete spectra in sample B. 
 Figures~\ref{fig:fighk} and \ref{fig:fighna} show examples of the fitting 
 of the atmospheric models, magnified around the Na~{\sc i}$\sim$8190 \AA \ and
 K~{\sc i}$\sim$7700 \AA \ doublet area for sample A.
 The minimum value of $S$-function for each target with M4-M8 spectral type is shown in 
 Tables~\ref{tab:resulta},~\ref{tab:resultb},~\ref{tab:resultab} and~\ref{tab:resultbb}.

\subsection{Analysis of objects of M8.5-M9.5 spectral type} 

 The atmospheric structure of objects of M8.5-M9.5 spectral type differs greatly from early M dwarfs. 
 The dust has a significant impact on the distribution of temperature vs. pressure in these atmospheres.
 Thus, it is incorrect to use the NextGen model for objects with temperatures T$_{\rm eff}\leq$ 2500 K. 
 We used the DUSTY and COND (Allard et al. 2001) atmospheric models for the study of the 
 late M dwarfs' synthetic spectra. The DUSTY and COND models were modified
 in an identical way to the model improvements for early M dwarfs: 
  we introduced additional scattering of radiation by dust clouds and reduced the 
 abundance of the molecules TiO and VO.

 As with earlier spectral types, we utilised fits to the TiO and VO absorption bands and the
 K~{\sc i} and Na~{\sc i} doublets, to determine the stellar fundamental
 parameters. In weakly ionized atmospheres at T$_{\rm eff}$$<$ 3000 K the natural broadening is
 several orders of magnitude weaker than pressure and Stark broadening, and can thus be neglected.
 The broadening of K~{\sc i} and Na~{\sc i} resonance lines is dominated by pressure effects,
 but the theory of this is not particularly advanced.
  Pressure broadening is calculated using two methods: a collisional approximation (van der Waals theory)
 and quasi-static theory (see Burrows \& Volobuyev 2003; Allard et al. 2003). The use of a collision
 approximation allows us to describe the profiles of metal lines in early M dwarfs.
 The quasi-static theory is used in the study of L dwarfs, where energy levels of  atomic sodium or
 potassium are immersed in a sea of molecular hydrogen and are subsequently perturbed by the potential
 field of the diatomic hydrogen. The modelling of atomic-line broadening for objects that are
 close to the transition between M and L spectral classes is complicated and this issue is
 beyond this paper aims. Since in our case we could see only the core of the lines and do not
 see extended wings, our fits should be sufficient without including
 any broadening.

 Physical properties were derived for objects with M8.5-M9.5 spectral types based on the fitting
 of high or low resolution spectra and synthetic spectra calculated for COND and DUSTY models.
 In addition $S$-function analysis provided the best fits. Since in sample B there are some objects
 classified as M8, we applied fits with the synthetic spectra (NextGen, DUSTY and COND)
 as well as with appropriate semi-empirical models, to provide the transition between the NextGen
 and the DUSTY/COND models.
 Results of these fits are shown in Tables~\ref{tab:resultab},~\ref{tab:resultbb},~\ref{tab:resultac},
  and~\ref{tab:resultbc}.

 LP944-20:
 LP944-20 is a known BD of spectral type M9 and we used it as a test for the atmospheric fits
 as well as to benchmark the age constraints described in following sections.
 Several authors have studied this object
 in detail to determine its characteristics. Tinney (1998) estimated an age between 475 and 650 Myr
 by evolutionary tracks, while Ribas (2003) determined LP944-20 as a kinematic member of the Castor
 moving group and therefore estimated an age of 320$\pm$80 Myr.
 Pavlenko et al. (2007a) used semi-empirical atmospheric models to determine the parameters of LP944-20
 including lithium abundance, of which the best fit gave log N(Li)=3.25 $\pm$ 0.25, and found
 a similar age to the one found by Ribas (2003).

 According to previous studies the effective temperature of LP944-20 is in the range
 2040 $\leq$ T$_{\rm eff}$ $\leq$ 2400 K (see Basri et al. 2000; Dahn et al. 2002; Pavlenko et al. 2007a).
 The best fit obtained here, with the DUSTY model, provides T$_{\rm eff}$=2400 K and log~$g$= 4.0
 (see Table~\ref{tab:resultac}).
 The log~$g$ obtained by Pavlenko et al. (2007a) was 4.5, with temperatures of 2000-2200 K
 using COND models.
 They remarked that temperatures in the outermost layers of the DUSTY models were higher in
 comparison to the COND models. This difference is consistent within the uncertainties.

\subsection{Results}

All the targets in sample A, with the exception of 2MASS0334-2130, present a better fit when
 a semi-empirical model with dust is used instead of the dust-free NextGen model,
 as shown by $S_{min}$ of Table~\ref{tab:resulta}.
 This would imply that the presence of dust should be taken into account for
 effective temperatures up to $\approx$2900
 (Jones \& Tsuji 1997, Kuznetsov et al. 2012).

 Similarly to LP944-20, the two targets of sample A cooler than M8 spectral type
 are better fitted by the DUSTY model (Tables~\ref{tab:resultab} and~\ref{tab:resultac}, 
 and Figures~\ref{fig:fight},~\ref{fig:fighk} and~\ref{fig:fighna}).

 For the analysis, we divided sample B into three different spectral type ranges for the fits:
 $<$M8, $>$M8 and as we have many objects with an M8
 spectral classification we treated them as a third set and applied both sets
 of models for them (see Section 4.4.).

 We used GL 406 (M6.0) as reference star to test spectral typing and model fits.
 However, we noticed that the flux calibration is very poor for $\lambda$$<$7500 \AA.
 The cause is uncertain since all other spectra were calibrated correctly.
 The K~{\sc i} and Na~{\sc i} lines were not present in the fault area, hence we centered
 our fit on $\lambda$$>$7500 \AA.

 Results of objects that were spectroscopically classified as earlier than M8 are shown in
 Table~\ref{tab:resultb}, where $S_{min}$ is given in columns 8 and 12.
 From these targets, four present a better fit with a semi-empirical model with dust,
 and two, 2MASS1236-6536 and 2MASS1736-0407, with NextGen. These two, were included in our sample
 due to their ``cold photometric" classification (Section 3.2) but spectral classification and model fit
 suggest warmer temperatures.

 For objects classified as M8, we applied all models. Results are given in
  Table~\ref{tab:resultbb}.
 In the case of 2MASS0814-4020, DUSTY and COND models have the same $S_{min}$, but the DUSTY
 model is more appropriate for the target's temperature. Also, despite the higher value of $S_{min}$
 for the NextGen model, we could appreciate by visual inspection and by the $S$-function 
 value that in parts of the spectrum NextGen was the best fit.
 For 2MASS1557-4350, the DUSTY and COND models also have the same $S_{min}$. We chose the DUSTY
 model as being more appropriate for its temperature.

 For objects with spectral types later than M8, parameters determined from the best fit to
 the observed SEDs are given in Table~\ref{tab:resultbc}.
 In this case, it was not easy to find a good fit for
 the blue part of the spectrum ($\lambda$$<$7500 \AA).
 In the case of 2MASS1909-1937 no COND model provided an acceptable result
 and the DUSTY model gave no reliable gravity values.
 It is likely that these objects are also suffering from relatively uncertain flux calibration, but
 the fact that these are the coolest objects and around the M-L transition, may indicate other issues
 like the use of an inapropriate set of models.
 Since the K~{\sc i} and Na~{\sc i} spectral regions seem to be correctly fitted we
 keep the results but consider them with caution for 2MASS1734-1151 and 2MASS1745-1640,
 while we could not extract any acceptable conclusion from the fit for 2MASS1909-1937
 (see Table~\ref{tab:resultbc}).

\section{Age characteristic}

  Sample A is formed of objects that are already confirmed as kinematic candidates
 of MGs. Since kinematics is not sufficent to confirm membership, determination of a
  candidate's age can further constrain its group membership to
 a point of robustness or dismiss it as an old field target that shares kinematics with a
 young MG. For our candidates, age in some cases can also discriminate between multiple MG candidatures.
 Sample B is formed of astrometric candidates and so further
 study should be applied to confirm any membership. This can be achieved by: 1)
 using gravity sensitive spectroscopic features to distinguish between young and old ultracool dwarfs,
 2) using the activity/age relation for late-type stars up to a spectral type of M7 (Mochnacki
 et al. 2002; Silvestri et al. 2006; Reiners \& Basri 2008; Jenkins et al. 2009),
  3) measuring the lithium 6708 \AA \ doublet (e.g. Rebolo et al. 1996; Pavlenko et al. 2007a)
 as the abundance of lithium is related to mass and age for substellar objects
 and 4) determining $v\sin{i}$ to differentiate between young and older M types.

\begin{figure*}
   \centering
   \includegraphics[width=7.0cm,angle=270,clip]{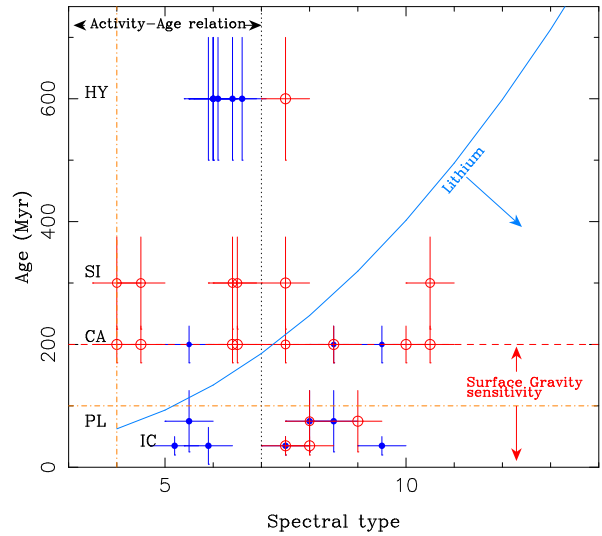}
      \caption{
 This figure (updated Figure~9 from Paper I), illustrates how an object would
 be selected for the various methods of age constraining. Candidates that appear
 to the right of the lithium edge (blue continuous line) can be followed up with a lithium test programme. 
 Objects that appear younger than 200 Myr (horizontal red dashed line), are eligible for
 follow-up using spectroscopic gravity sensitive features. Vertical and horizontal orange 
 dot-dashed line represents the Schlieder et al. (2012) limits for the Na~{\sc i} diagnostic
 applicability. We here took into account the 200 Myr limit. Candidates that fall to the left
 left of the spectral type=M7 limit (vertical black dotted line), would thus be
  suitable for age/activity relation follow-up, although candidates with a spectral type 
 close to M7 may be subject to large
 uncertainties on their age. Some candidate cannot be tested by any of these methods but
 will be eligible for age testing using $v\sin{i}$.
 Sample A targets are plot as blue circles and sample B as red circles. We plotted up to two
 candidatures for targets with multiple MG.}
         \label{fig:modelo}
   \end{figure*}

 Figure~9 of Paper I shows a plot of evolution-time (based upon potential group membership) versus spectral
 type for the complete sample of candidates. The figure, reproduced and updated here as
 Fig.~\ref{fig:modelo},
 illustrates how an object would be selected for various methods of age constraining follow-up mentioned.
 Candidates that appear to the right of the lithium edge can be followed up with a lithium test programme.
  Objects that appear younger than 200 Myr, are eligible for
 follow-up using spectroscopic gravity sensitive features. Candidates that fall to the
 left of the spectral type M7,  would thus be suitable for age/activity relation
 follow-up, although candidates with a spectral type close to M7 may be subject to large
 uncertainties on their age. Some candidates cannot be tested by any of these methods but
 will be eligible for age testing using $v\sin{i}$.

 In Paper II we presented a $v\sin{i}$ study (for sample A) where we applied this criterion of youth
 based on Figures~9 and 10 of Reiners \& Basri (2008) to support group membership
 (see Paper II). We study here the rest of the age-constraining features.

\subsection{Surface gravity features}

 The radius of young UCDs can be as much as 3 times greater than their eventual equilibrium
 state (Burrows et al. 2001), and as a result young objects can exhibit
 significantly lower surface gravities (10-100 times) than their evolved counterparts
 with the same spectral type. Low-resolution (e.g. R$\sim$350-2000) studies in the optical
(Mart\'in et al. 2010; Mart\'in et al. 1996; Luhman et al. 1997) and infrared
 (Jones et al. 1996: Gorlova et al. 2004; McGovern et al. 2004)
 have demonstrated that numerous features (e.g. CaH, K~{\sc i}, Na~{\sc i}, VO, etc) can be
 used as gravity-sensitive (and thus age sensitive) diagnostics for young objects.
 Diagnostics should be sensitive to gravity and age for objects that are
 younger than $\sim$200 Myrs (e.g. Barrado y Navascu\'es 2006). Schlieder et al. (2012) presented a
 study of the Na~{\sc i} doublet equivalent width in giants, old dwarfs,
 young dwarfs, and candidate members of the $\beta$ Pic moving group using medium
 resolution spectra. They concluded that the diagnostic is reliable for objects with
 spectral types later than M4 and younger than 100 Myr, and that metallicity has
 an important role. Therefore, this youth indicator is best used on samples
 with similar metallicity. Thus, we aimed to apply this test by determining the
 surface gravity of the younger MG candidates, although we took into account all
 candidates up to 200 Myr.

 As mentioned before, the main atomic gravity-sensitive features present in our spectra are:
 the K~{\sc i} resonance doublet (7665 and 7669 \AA) and
 the Na~{\sc i} subordinate doublet (8183 and 8199 \AA).
 We fitted the spectra of the candidates with a synthetic model (see Section 4)
 to determine the log~$g$ of both samples A and B (Table~\ref{tab:resulta},
 ~\ref{tab:resultb}, ~\ref{tab:resultab}, ~\ref{tab:resultbb},~\ref{tab:resultac}, and ~\ref{tab:resultbc}).
 Figures~\ref{fig:fighk} and \ref{fig:fighna} show some examples of atmospheric model fitting
 magnified on the aforementioned features for sample A.

 We performed two approaches in the surface gravity studies: inferring age-limits from
 1) the surface gravity obtained from the best fit synthetic models
 and 2) comparison of the equivalent width of gravity-sensitive Na~{\sc i} doublet.

\subsubsection{Values of $log~g$ from synthetic model fits}

 As mentioned above, surface gravity features can be studied to discriminate if an
 object is young or old, but it is not easy to associate a log~$g$  value to an
age interval, that is, to have an absolute log~$g$-age relation.
 Many parameters affect the derived log~$g$: atmospheric model used,
  resolution and instrument.
 Therefore, we observed LP944-20, with known age and previously fitted by different models, to
 set the range of log~$g$-age interval for our observations and as a test
 object for our method of late M dwarf investigation.

  Taking into account the association to the Castor MG and the age based on the presence of
 lithium, we used our log~$g$ for LP944-20 as a reference and compared it to
 the log~$g$ of the candidates. From this, we assume that an object with spectral type
 and log~$g$ similar to LP944-20 is probably of similar age.
 Models from the literature (e.g., NextGen, DUSTY, COND), associate similar or lower
 ages with warmer objects with the same log~$g$ than to those that are cooler.
 For example, selecting log~$g$ in the 4.0-4.5 range in DUSTY and COND models,
 an object with T$_{eff}$ $=$ 2400 K intersects 8-30 Myr age isochrones, whilst
 objects with T$_{eff}$ $\geq$ 2400 K intersect 5-7 Myr, and objects with
 T$_{eff}$ $\leq$2400 K intersect with isochrones over 40 Myrs.
 Both the DUSTY and COND models present ambiguities for 30 Myr isochrones. The COND model
 presents similar problems for $\geq$40 Myr isochrones, but fortunately in this case with
 temperature ranges far from that of our targets.
 Therefore, we can assume that objects warmer than LP944-20 would have, in
 general, ages $\leq$200-300 Myr should they have log~$g$ = 4.0.

 Since LP944-20 is only a single object, in order to constraint this log~$g$-age relation, we
 also used the GL 876 field target, observed with the candidate A sample.
 This allows us to compare our target log~$g$ values with those of old and
 intermediate age targets.

 We fit them in the same way as our candidates and results of the best fit are given in
 Table~\ref{tab:resulta} and~\ref{tab:resultac}.

 As mentioned in Section 4.4, LP944-20 is a known member of the Castor
 MG (age$\sim$200 Myrs) and Pavlenko et al. (2007a) obtained an age of
 $\sim$300 Myr through lithium depletion. Applying the atmospheric model
  method described in Section 4, we obtained a log~$g$ of 4.0$\pm$0.5.

 GL 876 is an M4 dwarf. It is known as an exoplanet host
 of four planets. Literature provides few age calculations, such as 1-10 Gyr
 from Marcy et al (1998), $>$1 Gyr from Shankland et al. (2006) and 0.5-1 Gyr
  from Correia et al. (2010). Despite the variations it is clearly an old object that
 should show a high value of gravity. We find that the NextGen model best fit
 provides a log~$g$ of 5.5$\pm$0.5.

To set the age-gravity relation for our sample, we used these results as a basis.
   Log~$g$ obtained values of the candidates are 4.0, 4.5, 5.0, and 5.5, where we can state that
 objects with log~$g$=4 are young and still in contraction and that objects with
  log~$g$=5.5 are old. With a 0.5 uncertainly, an object with log~$g$ of 4.5 (4.0-5.0)
 shows ambiguity when discriminating between young and old.

  Accepting the limitations and the uncertainty in the constraints, spectral type and metallicity
 dependence, etc, for the atmospheric fits to our sample, 
 objects with log~$g$$\approx$4.0 are probably young
 (under $\sim$200 Myrs), objects with log~$g$$\approx$4.5 are older but possibly still in contraction
 ($\sim$300 Myrs), and therefore objects with log~$g$$\approx$5.0 and log~$g$$\approx$5.5
  are probably older than 300 Myrs.

 The log~$g$ obtained with model comparison are given in columns 5 and 9 of
 Table~\ref{tab:resulta} and~\ref{tab:resultac}, columns 6 and 10 of
 Table~\ref{tab:resultb} and~\ref{tab:resultbc},
 and columns 5, 8 and 11 of Table~\ref{tab:resultab} and~\ref{tab:resultbb}.

\subsubsection{Na~{\sc i} equivalent width ($EW$) comparison}

 Additionally, we checked if $EW$(Na~{\sc i}) measurements can be used to determine youth, based
 on Mart\'in et al. (2010) and Schlieder et al. (2012).

 Given the cool nature of the targets, the measurement of equivalent widths in the optical are
  generally measured relative to the observed local pseudo-continuum formed by molecular
 absorptions (mainly TiO), and therefore the $EW$s are ''pseudo-equivalent widths"
 (Pavlenko 1997; Zapatero Osorio et al. 2002). However, we will call them $EW$ for simplicity
 hereafter.

\begin{figure*}
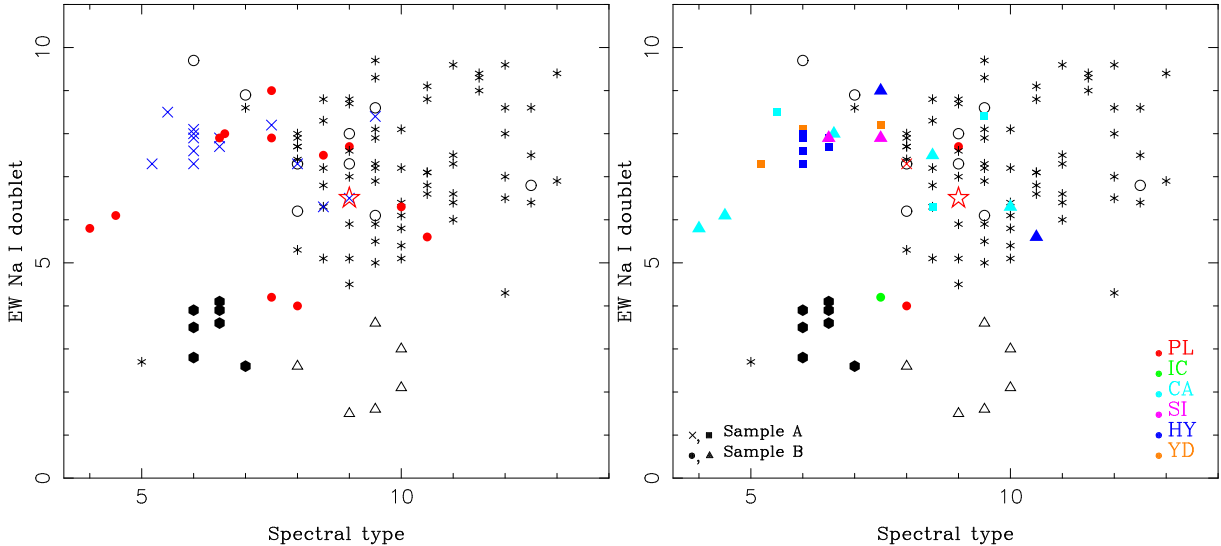

   \centering
   \includegraphics[width=8.0cm,clip]{Figura8n.ps}
   \includegraphics[width=8.0cm,clip]{Figura8nb.ps}
     \caption{
  The right and left panels show two versions of the same information with different legends in order to highligh different 
 information.
  Left: From Mart\'in et al. (2010), 65 high-gravity field objects are
  plotted as six pointed asterisk symbol, 6 low-gravity objects as open triangles,
 12 reference field stars as open circles and 7 Upper Sco candidates as solid hexagons.
  We overplot sample A as blue crosses, and sample B as red filled circles. 
 The red star marks LP944-20. Errors are approximately the size of plot symbols. 
 Right: Here we overplot to Mart\'in et al. (2010) data, sample A (crosses and filled squares),
  and sample B (filled circles and triangles), plotted in different colours depending on their
 final MG membership candidature (Table~\ref{tab:resultfinal}). When objects present more than one candidature,
  we plotted the membership to the youngest MG. 
 To better discriminate young targets, we use triangles and squares for candidates finally 
 classified as possible member of MGs with ages $\geq$200 Myr while crosses and circles are candidates
 finally classified as possible member of MGs with ages $<$200 Myr (see Sect. 6).
          }
         \label{fig:ewnacomp}
   \end{figure*}

 Although there is a considerable scatter in the data (mainly from the use of different instruments)
  Mart\'in et al. (2010) Figure~3 plots Na~{\sc i} $EW$ versus spectral class for a few
 field objects and members or candidates of the Upper Scorpii OB association.
 To measure consistent equivalent widths in spectra of different resolutions, they established
 as a rule that the pseudo-continuum region was between 823 nm and 827 nm, and integrated the
 line from 817.5 nm to 821.0 nm.
 Their figure shows a clear trend, and they inferred that objects with the weakest Na~{\sc i}
 are likely to have low surface gravity. They established as a rule of thumb that any object with
 spectral class between M6 and L4 and with a Na~{\sc i} doublet detectable but weaker than
 field counterparts observed with the same spectral resolution, is likely to have a low
 surface gravity and consequently a very young age (i.e., younger than 100 Myr) and
  substellar mass.
  Sample B have similar resolution to the Mart\'in et al. (2010) data. As such we
 measured the combined equivalent widths of the two lines of the Na~{\sc i} doublet
 in the same manner (given in Table~\ref{tab:resultfinal}) allowing direct comparison.
 In the case of sample A, we convolved the spectra to the same resolution and then
  measured $EW$(Na~{\sc i}).
 The left panel of Figure~\ref{fig:ewnacomp} presents equivalent widths of the Na~{\sc i}
 doublet versus spectral type
 for all targets of sample A as blue crosses, and B as red filled circles,
  with the Mart\'in et al. (2010) data: 65 high-gravity field objects are
  plotted as six pointed asterisks, 6 low-gravity objects as open triangles,
 12 reference field stars as open circles and 7 Upper Sco candidates as solid hexagons.

 While the metallicities of all MGs are similar to solar, the Upper Sco cluster does
 not yet have a robust metallicity determination, and other objects in
 Mart\'in et al. (2010) have different metallicities.
 We took the information from $EW$(Na~{\sc i}) studies
 in Schlieder et al. (2012) into account when compare
 samples A and B with Mart\'in et al. (2010).

 For sample A, if we look into moving group candidatures (thus age) and $EW$s(Na~{\sc i}),
 they are inconsistent (Table~\ref{tab:resultfinal} column 2). That is to say that some
 young MG candidates present a higher $EW$(Na~{\sc i}) than others which have candidature to
 older MGs. But if we take into account the final MG membership given
 in column 11 of Table~\ref{tab:resultfinal}, most of sample A targets are finally classified
 as candidates to MGs with ages $\geq\sim$200 Myr (except one, SIPS2039-1126),
 and therefore the $EW$(Na~{\sc i}) become consistent (see Figure~\ref{fig:ewnacomp} right panel).
 SIPS0007-2458 and 2MASS0334-2130, both candidates to IC2391 present values of $EW$(Na~{\sc i})
 too high for IC2391 age. Since we obtained $EW$(Na~{\sc i})=6.5 for LP944-20,
 the value of $EW$(Na~{\sc i})=7.3 for SIPS2039-1126 is comparatively high for its Pleiades membership.
 Convolution might have introduced additional noise affecting the comparison and we discuss
 further this target in Section 5.3.

 For sample B, if we look into moving group candidatures and $EW$s(Na~{\sc i})
 (Table~\ref{tab:resultfinal} column 2), they are consistent.
 That is in general, target candidates to IC2391 and the Pleiades show the lowest $EW$s, candidates to
 Castor and Sirius show higher, intermediate $EW$ values and candidates to Hyades show
 the highest values of $EW$ (Figure~\ref{fig:ewnacomp} right panel).
 The two targets showing low values of $EW$s(Na~{\sc i}) are 2MASS1326-5022, candidate to IC 2391
 and 2MASS1557-4350, candidate to Pleiades and IC 2391. The rest of the candidates
 show higher values in agreement with their MG candidatures except for 2MASS1734-1151, M9 candidate to
 Pleiades but showing larger values of $EW$s (Na~{\sc i}) than LP944-20.

\subsection {Activity vs. age relation}

 Subsequent studies by Skumanich (1972) show that activity
 decreases over time for late-type stars. However, after a spectral type of $\approx$ M3-4,
 stars become fully convective and the activity is driven by a turbulent
 dynamo in a mechanism that is still not clear.
 Many studies have found evidence that the age-activity relation extends
 into the M dwarf regime (see e.g, Silvestri et al. 2006; 
 Mochnacki et al. 2002; Mohanty \& Basri (2003); Reiners \& Basri 2008).
 Recently, some authors (Berger et al. 2008; West et al. 2008) have shown that
 the fraction of active M0-M7 stars decreases with the vertical distance from the Galactic plane,
 further evidence of an age-activity relation.
 Comparing activity data to dynamical simulations, West et al. (2008) 
 derive an "activity life-time" relation for M dwarfs of spectral type M0-M7 in the range 0.4-8 Gyr.
 Following this, we can identify older non-MG UCDs in our sample through lack of 
 chromospheric emission.

 Comparing activity levels of our sample with activity levels of objects of the same spectral
 type and with known age, we can infer age ranges for our targets that would help us
 to confirm (or otherwise) their MG membership in the same way of other literature examples such us
 Stauffer et al. (1995; Fig.~13), Stauffer et al. (1997; Fig.~8), Terndrup et al. (2000 ; Fig.~9), 
 Barrado y Navascu\'es et al. (2004; Fig.~5), Shkolnik et al. (2009; Fig.~13).

 Although there are several chromospheric activity indicators (Ca~{\sc ii} H\&K, Balmer lines, 
 Ca~{\sc ii} infrared), we used the H$\alpha$ emission since it is included in the spectra presented here
 (with the exception of the SIPS2039-1126 spectrum).
 Young M dwarfs generally have H$\alpha$ in emission, with the average equivalent width increasing 
 to later spectral types (e.g. West et al. 2004). 
 Generally the level of chromospheric activity is measured by comparison of the diagnostic line's flux, but
  for M dwarfs equivalent width is also used. 
 We derived H$\alpha$ pseudo-equivalent widths by direct integration of the line profile using the 
 {\sc splot} task in IRAF, however, we will utilise them as $EW$s.

 Taking into account that $EW$(H$\alpha$) will reflect any activity variability, often present
 in young objects, and that moving groups with different ages will show different activity 
 saturation levels as well as a higher dispersion in the level of activity when they are older 
 (see e.g. Stauffer et al. 1997, Pizzolato et al. 2003; Ryan et al. 2005; L\'opez-Santiago 2005;
 Mart\'inez-Arna\'iz et al. 2011),
  we compared the $EW$(H$\alpha$) versus spectral type of our candidates with 
 objects of known age (obtained from other methods). 
 
 We compiled a sample of M dwarfs with known age and $EW$(H$\alpha$) from the literature.  
 These were taken from Hyades and Pleaides members in Terndrup et al. (2000), young and old 
 disk and field stars in Mohanty \& Basri (2003) and from a compilation of targets with ages determined 
 by different methods in Shkolnik et al. (2009). When using targets with MG or cluster  memberships, we assigned 
 them the age of the cluster. 
 From Mohanty \& Basri (2003), we assigned an averaged value of 500 Myr to YD,
  600 Myr to YO (old-young), 1 Gyr for OD and more than 1200 Myr for Halo targets. 

\begin{figure*}
   \centering
   \includegraphics[width=8.5cm,clip]{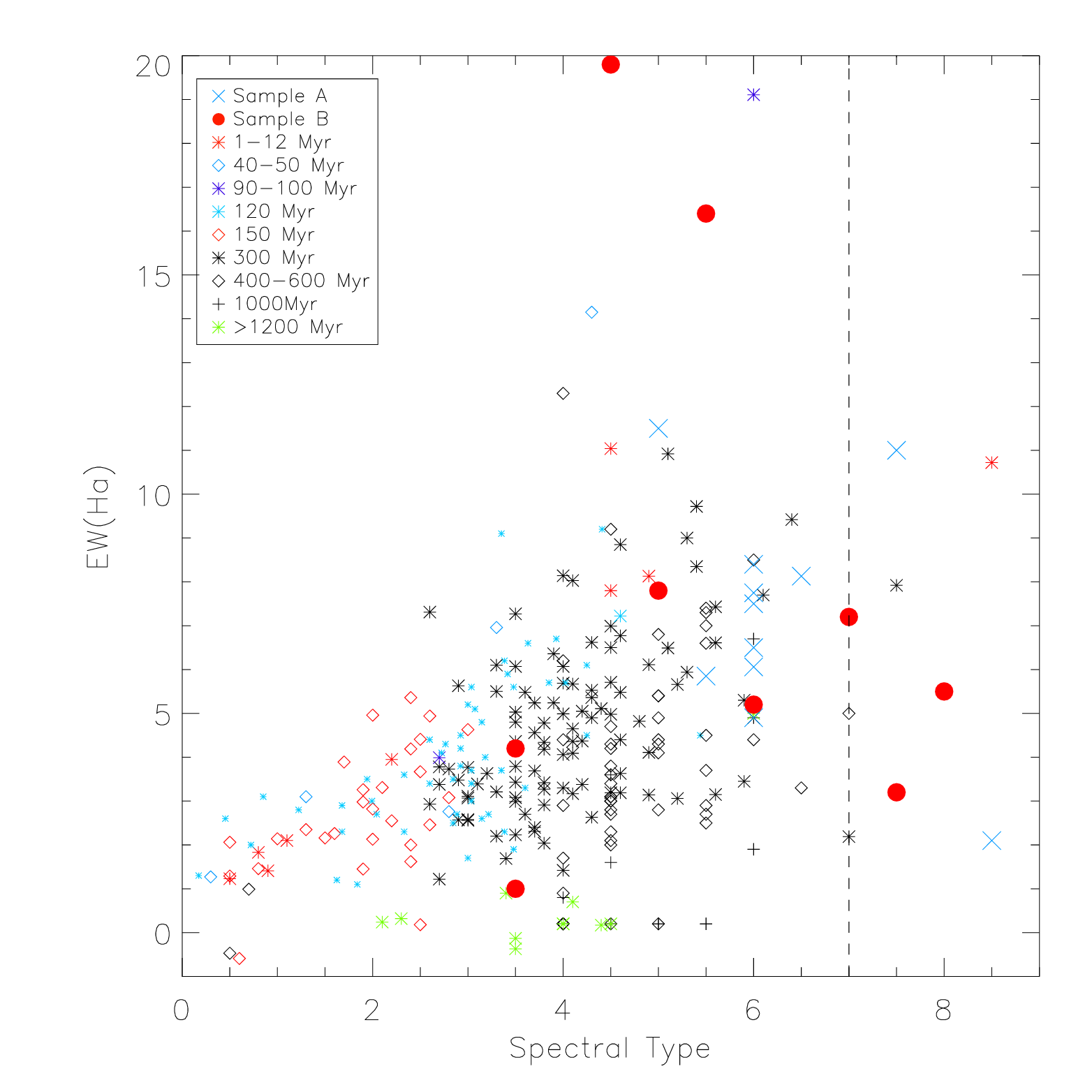}
   \includegraphics[width=8.5cm,clip]{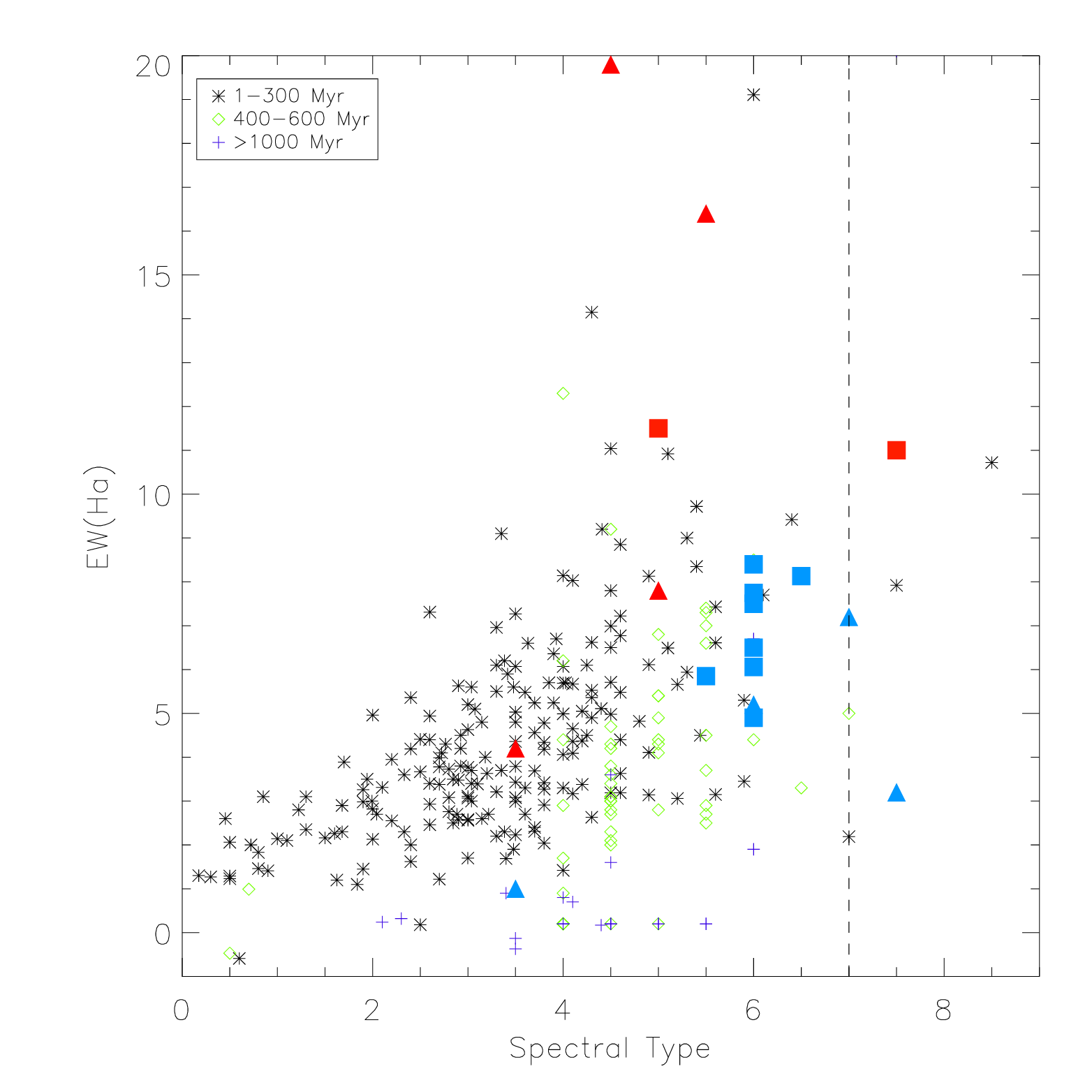}
     \caption{
  $EW$(H$\alpha$) versus spectral type. In the same way as in Fig.~7, both panels show two versions of the same 
 information with different legends in order to highligh different information.
 Left: Sample A is plotted as blue crosses, Sample B as red circles and literature known-age 
 objects in different symbols according to age: 
 red asterisk for objects with age between 1-12 Myr, blue diamonds for
 40-50 Myr, violet asterisk for 90-100 Myr, blue asterisk for 120 Myr, red diamonds for
 150 Myr, black asterisk for 300 Myr, black diamonds for 400 Myr and green asterisk for
 objects with ages over 1200 Myr. Literature data has been obtained from  
 Terndrup et al. (2000), Mohanty \& Basri (2003) and Shkolnik et al. (2009) 
 where ages were calculated by known-age MG or Cluster
 membership or by other age constraining methods.
 Right: Similar to left panel where to favor the age discrimination we 
 plot only three age intervals. 
 Sample A targets are plotted as filled squares and sample B as filled triangles,
 where targets classified with ages $\geq$300 Myr are plotted in blue and
 targets with ages $<$300 Myr are plotted in red.
 }
         \label{fig:ewhacomp}
   \end{figure*}

 The $EW$(H$\alpha$) of our objects is given in column 3 of Table~\ref{tab:resultfinal}.
 Figure~\ref{fig:ewhacomp} left panel presents $EW$(H$\alpha$) versus spectral type 
 for our candidates, with blue crosses and red filled circles for sample A and B respectively, 
 in comparison with data from the literature.

 In the right panel of Figure~\ref{fig:ewhacomp}, only three ranges of ages are plotted to
 better discriminate the possible age interval of our samples. 
 Sample A targets are plotted as filled squares and sample B as filled triangles,
 where targets classified with ages $\geq$300 Myr are plotted in blue and
 targets with ages $<$300 Myr are plotted in red.

 Although theoretically equal, equivalent widths measured in lower resolution spectra can be
 slightly larger than measured in higher resolution spectra (due to line blending),
 and pseudo-equivalent width measurements show variation with resolution, therefore
 there is a possibility that variability effects are seen in our observations.
 Taking into consideration that only one data point is available for each of our candidates (and therefore 
 variability can not be determined), that older MG members can present higher scatter, and that
 for sample B, values of $EW$(H$\alpha$) have been measured at low-resolution, we have determined
  a rough age interval for 18 of the candidates (we have included targets in the 
 limiting spectral type M7 and M7.5).

 We have assumed that the candidates are single objects. Any stellar or substellar
 companion would probably increase the activity levels making the targets look younger
 when we measure $EW$(H$\alpha$) (e.g. Basri et al. 1995; Ruten 1987; 
 Schrijver \& Zwaan 1991; G\'alvez-Ortiz 2005).

 All targets in sample A and B with spectral types under M7.5,
 have $EW$(H$\alpha$) values  inside
 the limits of an age that agree with at least one of their MG candidature (see columns 3 and 4 of
  Table~\ref{tab:resultfinal} and Figure~\ref{fig:ewhacomp}).

 It should be noted that the SIPS2039-1126 spectrum does not include the H$\alpha$ region, 
thus the activity criterion has not been used in this case.

\subsection{Lithium}
The Li~{\sc i} doublet at 6708 \AA \ is an important diagnostic of age in young
 late-type stars. Lithium is destroyed in fully convective low-mass objects,
 with mass from 0.3-0.06M$_{\odot}$ (e.g. Rebolo et al. 1996) as all acquire core temperatures 
  \hbox{$> 2 \times 10^6$ K} as they heat up during early contraction. 
 Convection ensures that atmospheric lithium abundance reflects
 core lithium content, and the observable lithium doublet thus acts as a gauge for
 the core temperature and contraction age. The so-called "lithium edge" 
 that separates objects with lithium from those without, can be seen in co-eval
  populations (e.g. in open clusters and moving groups), through the spectral type (or mass). 
 The lithium edge advances towards later spectral type as a population ages 
 (see Fig.~9 and caption of Paper I), and the presence
 or absence of lithium thus provides a critical age constraint as a function of an object's
 spectral type. 

 Due to the low resolution and the presence of artifacts in the Li~{\sc i} spectral region 
 for sample B, we searched for the Li~{\sc i} doublet only in sample A candidates.
 The Li~{\sc i} region in the sample A spectra show S/N values of no more than 10 in all the 12
 candidates which hampers the search for the absorption line.
 We found the Li~{\sc i} absorption line in SIPS2045-6332 and SIPS2039-1126 only, but 
 we do not discard the possibility of other fainter candidates with no signal in 
 this region, presenting some absorption when better spectra can be achieved.
 We also measured Li~{\sc i} in LP944-20 with the parameters found here, obtaining
 logN(Li) = 3.00 $\pm$ 0.5 dex, using synthetic spectra with T$_{\rm eff}$=2400 K, $log~g$ = 4.0, 
 DUSTY atmosphere structure and a rotational velocity of 35 km s$^{-1}$ (measured by us). This is
 within the uncertainties of other literature findings (see Section 4.4.1). 

 Clarke (2010) found the lithium signature in SIPS2045-6332 and SIPS2039-1126 suggesting
 youth and a BD nature inside the MG membership. Here, 
 we obtained the theoretical Li~{\sc i} pseudo-equivalent widths, relative to the computed pseudo-continuum formed 
 by molecular absorption, via direct integration of the line profile over the spectral interval 6703.0-6710.8 \AA.
 Figure~\ref{fig:lithium} shows the Li~{\sc i} area for SIPS2045-6332 and SIPS2039-1126 with their respective
 fits. Li abundance of logN(Li) = 3.5 $\pm$ 0.5 dex for SIPS2045-6332 and logN(Li) = 3.0 $\pm$ 0.5 dex 
 for SIPS2039-1126 provides age estimations as young as 7-100 Myrs or the possible BD condition.

\subsection{New Brown Dwarfs}

 The lithium test was first proposed by Rebolo, Mart\'in \& Maggazz\'u (1992) and developed by
  Magazz\'u, Mart\'in \& Rebolo (1993) to distinguish BDs from stellar objects.
 While stars and low mass stars deplete lithium with time (see explanation in Section 5.3),
 substellar objects with M$<$0.06 M$\odot$ mass can not achieve the temperature needed to
 destroy lithium and so it should be preserved independently of the object's age
 (Chabrier \& Baraffe 2000; Basri et al. 2000).
 Since young low mass objects have not yet depleted all lithium,
 the discrimination between stars and BD through the lithium test should take age into account.

\begin{figure*}
   \centering
   \includegraphics[width=6.0cm,angle=270,clip]{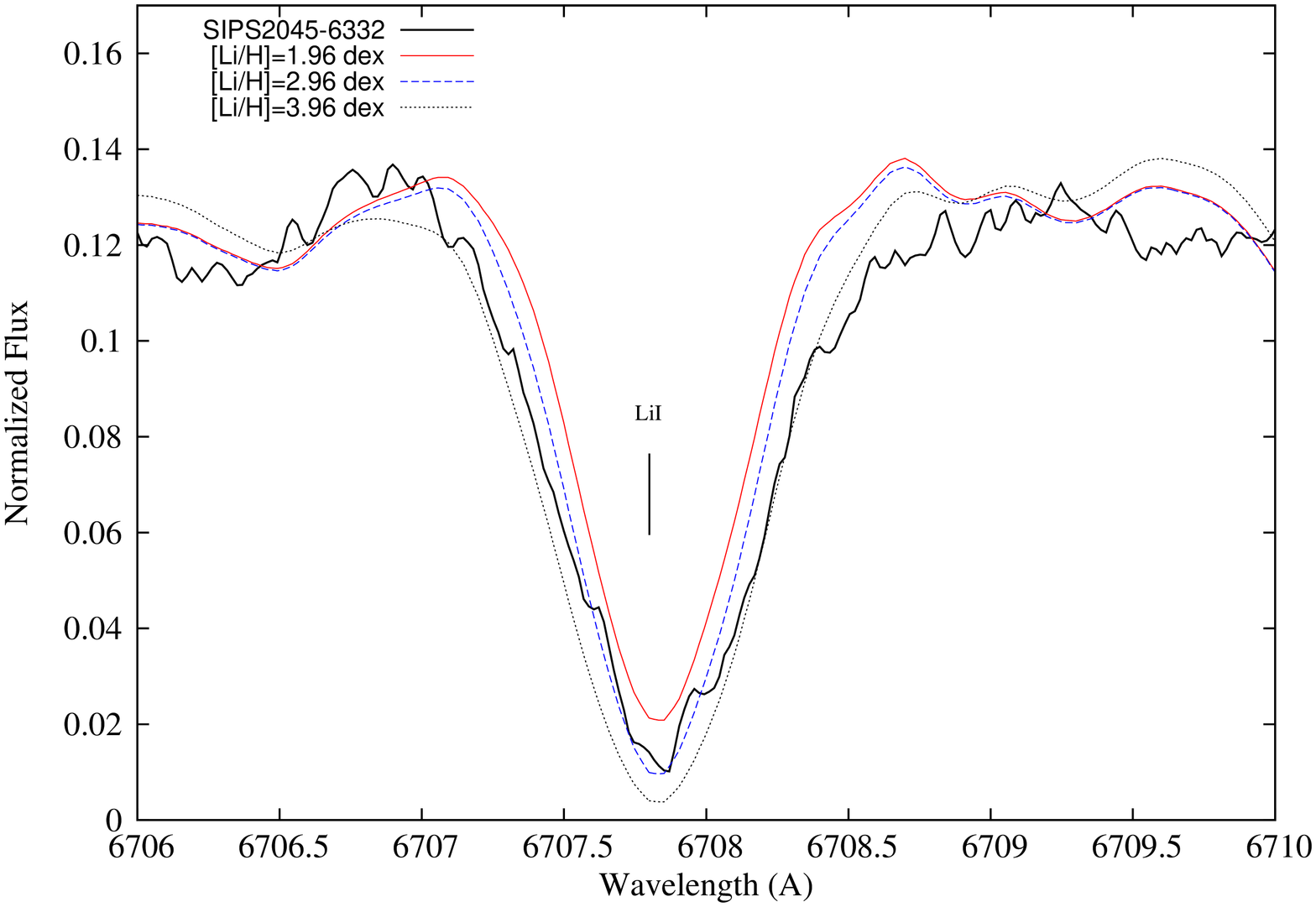}
   \includegraphics[width=6.0cm,angle=270,clip]{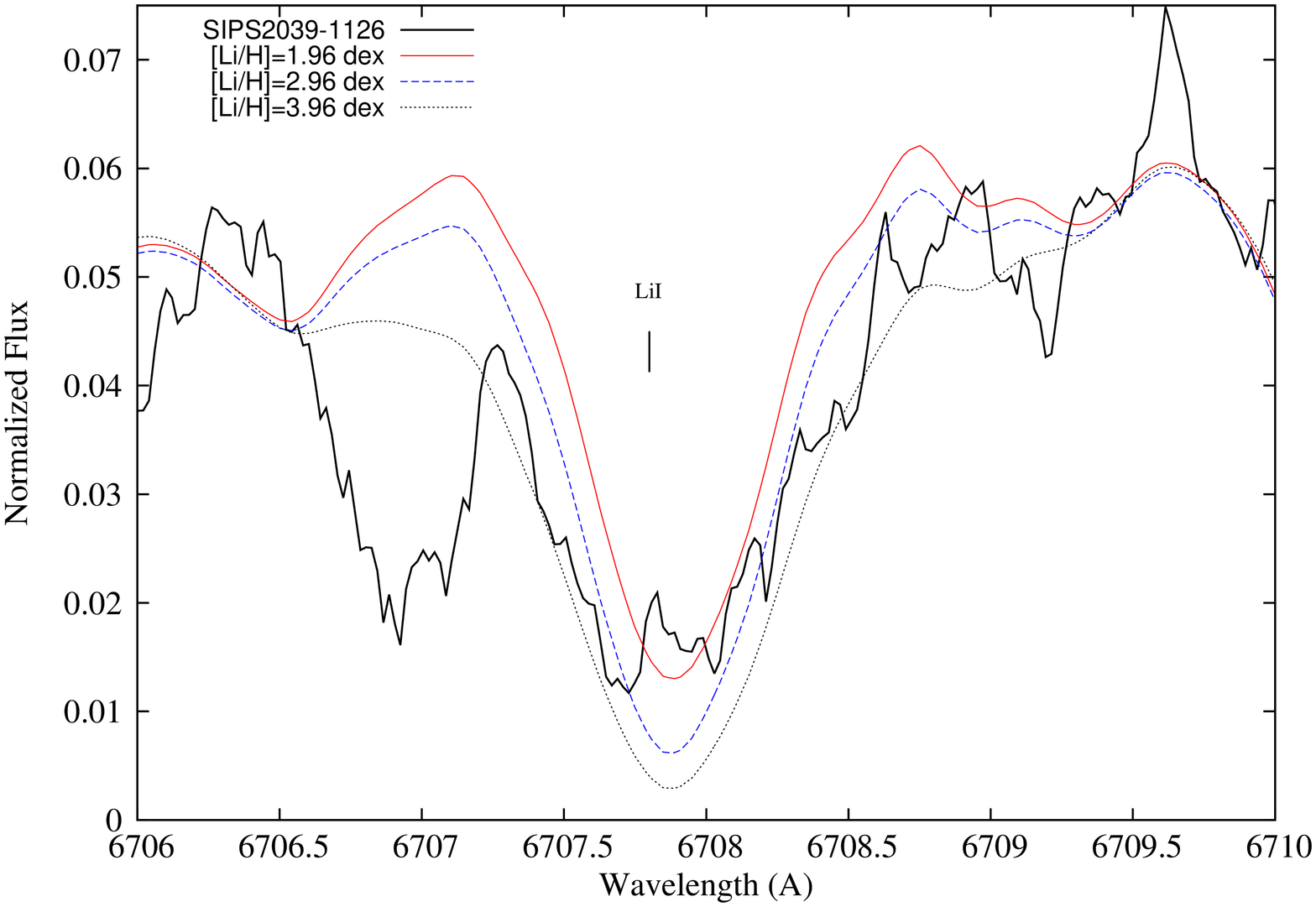}
     \caption{Fit to the observed spectra across Li~{\sc i} resonance doublet
 for SIPS2045-6332 (left) and SIPS2039-1126 (right).
 Best synthetic fits are overplot with different Li abundances.
              }
         \label{fig:lithium}
   \end{figure*}

\begin{itemize}
\item {\bf SIPS2045-6332:}
 SIPS2045-6332, was classified through several spectral indices as an M8.5 spectral type (paper II).
 It shows a clear Li~{\sc i} absorption line (Figure~\ref{fig:lithium}). 
 We used the parameters found in the best fit of the synthetic models and
 calculated its lithium abundance, logN(Li) = 3.5 $\pm$ 0.5 dex. The DUSTY best fit
 $log~g$ values of 4.0$\pm$0.5 and this lithium provide an age that confirms 
 SIPS2045-6332 as probable Castor member.

 We noticed that $EW$(H$\alpha$) value, 2.1, is quite small for a young object. But
 with a M8.5 spectral type, this value agrees with the findings of other authors,    
 e.g. Mohanty \& Basri (2003), Reiners \& Basri (2010), Barnes et al. (2013), 
 that found that the H$\alpha$ emission is roughly constant from mid- to late M,
  but there is a sharp drop between M8-L0 (see e.g. Fig.~5 of Mohanty \& Basri 2003), 
 from which they show little or significantly reduced emission in spite of
  significant rotation.

 At the Castor MG's age, and with a spectral type
 of M8.5 (M$<$0.06M$\odot$), lithium confirms it also as a BD.

\item {\bf SIPS2039-1126:}
 Similarly,  SIPS2039-1126,  was classified through several spectral indices as an
 M8.0 spectral type (paper II).
 It shows a Li~{\sc i} absorption line (Figure~\ref{fig:lithium}) although the S/N of 
 the area is only $\sim5$. Using the parameters from the best fit of the synthetic models, 
 we obtained logN(Li) = 3.0 $\pm$ 0.5 dex (Section 5.3).
 The DUSTY best fit $log~g$ values of 4.5$\pm$0.5 suggest it is probably still in contraction.
 In Section 5.1 we saw that $EW$(Na~{\sc i}) of SIPS2039-1126 was high considering 
 its Pleiades membership. We have used gravity criteria for targets up to 200 Myr but as seen 
 in Fig.~\ref{fig:modelo}, following Schlieder et al. (2012), some Pleiades members may
  lie outside these limits. 
 We thus can not discard this target as a PL member taking into
 account the rest of the evidence.

  At the Pleiades age or in the age interval suggested by the gravity values, the lithium 
 depletion boundary occurs at M6.5 (M=0.075M$\odot$; Barrado y Navascu\'es et al. 2004). 
 Thus, with an M8.0 spectral type, lithium confirms SIPS2039-1126 as a BD.

\end{itemize}

\section{Summary and conclusions}

 We have presented a study of the spectral signatures of 25 low-mass
  objects that were candidate members of five young moving groups.
 We studied different typical age-constraining spectroscopic criteria utilising high and low resolution
 spectra, and combined the results to extract a final membership assessment for each target.

 We took into account spectral classification and thus approximated atmospheric temperature to
 apply the most appropriate model or models in the search of the best physical parameter constraints.
 When we used the synthetic stellar atmospheric models, the good agreement between
 observed and theoretical energy distributions suggests that semi-empirical models describe well the
 impact of dust in the atmospheres of M dwarfs. $S$-function analysis shows that the inclusion
 of dust effects makes it possible to achieve better fits for objects with $T_{\rm eff}\leqslant2800$~K.

 In both samples, although youth can generally be established, $log~g$ and $EW$(Na~{\sc i})
 were not useful for discriminating between different MGs. Some targets of sample A that were
 candidates to the HY moving group having lower $log~g$ values than candidates
 to younger MGs, and similarly in $EW$(Na~{\sc i}) values. In sample B the values are more consistent
 but most of the targets have several candidatures, making the situation more complicated.
 However, looking at the final membership results, only one target of sample A is classified as
 a member of a MG with age $<$200 Myr, which makes this criterion non applicable for most of them.
 For sample B, in spite of the multiple membership candidatures in the final results, we
 can discriminate between the lower values of $log~g$ (or lower values of $EW$s(Na~{\sc i})) for
 targets that are members of younger MGs, medium values for targets that are members of intermediate age
 MGs and the higher values for targets that are members of oldest MGs, which finally is
 consistent with the $log~g$-age relation. The comparison of $EW$(Na~{\sc i})s
 is probably useful to discriminate very young targets (less than 100 Myr) in a homogeneous sample
  (as suggested by Schlieder et al. 2012) while $log~g$ values are probably useful to discriminate 
 very young (less than 100 Myr) and very old (more that 600 Myr) objects, 
 serving as complementary study to other criteria.

 The activity-age relation is a reasonably useful age criterion up to $\approx$M7 spectral type but
 activity variability, binarity, and the comparison of different instruments/resolutions can introduce
 considerable uncertainties in age estimation. The presence of Lithium discriminates between young and old
 low mass stars but an ambiguity between young low mass stars and BDs limits its use in this
 cool temperature region. Gravity sensitive features and rotational velocity can be
 useful youth indicators in support of activity and lithium diagnostics, but clearly 
 have large dispersion and they need to be applied in conjunction with similar data.
 For a reliable constraint of cool object ages, a combination of different criteria are needed.

 We find 10 of the 13 objects from sample A to be probable members of one of the MGs,
 and also that 2MASS0020-2346 is a young disk member.
 SIPS0007-2458, a candidate member of the IC2391 MG, shows positive candidature under most criteria but does
 not show the expected lithium (associated with the age of the IC2391 MG). Due to the age constraint from $H\alpha$
 emission and surface gravity, we are inclined to think that SIPS0007-2458 is probably a young object from the YD
  class, but is not in the IC2391 MG, although it could also be a contaminant field object.
 2MASS0334-2130 is a similar case, with an M4.5-6 spectral type, it should contain lithium if it were
 an IC2391 MG member. Except for the $H\alpha$ criterion, that could be reflecting a maximum in activity level or
 the influence of an unseen companion, everything indicates that 2MASS0334-2130 is older than the IC2391 MG.
 Thus we also conclude that 2MASS0334-2130 is not an IC2391 member, although it could be a young object
 from the YD, or a contaminant target from the field.

 In sample B, 10 objects could be MG members, although more information
 is needed to discriminate between the possible MG membership of objects with several candidatures.
 We could not obtain any results for 2MASS1909-1937, so this remains an astrometric
 candidate to the Castor MG. 2MASS1734-1151 gave contradictory results so it
 also remains an astrometric Pleiades MG candidate until further analysis can be performed.

 We also confirm two moving group candidates, SIPS2045-6332 and SIPS2039-1126, as BDs.

 With all the acquired information, we find that 85\%, 83\% and 84\% of samples A, B, and both combined,
 show spectroscopic signatures of youth in agreement with the age of the moving group to which they present
 kinematic membership. This result suggests that there is a fairly low rate of contamination in such
 kinematic candidate samples. Also, there are candidates that cannot be confirmed or dismissed with the
 information obtained, which might (in the future) increase the final confirmation rate.
 We note that additional diagnostics to assess membership may be measured in the future, such as
 chemical tagging. Recently, in a kinematical-chemical investigation of the AB Dor moving group “Stream”,
  Barenfeld et al. (2013) shows that kinematics, color-magnitude positions, and stellar youth indicators
 alone can be insufficient for testing whether a kinematic group of stars actually shares a common origin.
 Future chemical study of our MG candidates will complement this study.

 In addition, this study has also helped to test and improve the atmosphere models for cool dwarfs
 (see Kuznetsov et al. 2012 and Kuznetsov et al. 2013). Youth can be an advantage
 for various reasons. Because young objects are brighter they can be detected and studied out 
 to greater distance than older, fainter objects.
 Also, the proximity of the MGs can allows the exploration of the
 faint circumstellar environment at relatively small distances from the star. Indeed, we will target our
 strongest MG candidates to search for lower mass companions using high resolution imaging techniques (e.g.
 adaptive optics). The discovery of planetary systems around young and low-mass objects provides crucial
 information for the understanding of planetary and stellar formation.

 From our complete study of the 80 objects considered (within papers I, II and this paper),
 we have found a total of 45 new possible MG members (with 8 of them showing more than one MG candidature)
  and around 21 possible young disk objects with no clear membership of the 5 MGs considered.

 Of these we find 26, 13, 6, 4 and 5 possible members of the Hyades, Castor, Ursa Major, Pleiades and IC2391 MGs
 respectively. Tables~\ref{tab:comptod} and \ref{tab:compothers} of the appendix compile all the M-L dwarfs
 investigated in our study.

\begin{table*}
\caption[]{A: Properties for sample A  
\label{tab:samplea}}
\begin{flushleft}
\begin{center}
\small
\begin{tabular}{lllccc}
\noalign{\smallskip}
\hline
\noalign{\smallskip}
Name & $\alpha$ (2000) & $\delta$ (2000) & J-mag & Moving group$^{1}$ & $V\sin{i}$$^{2}$  \\
  & (h m s) & ($^{\rm o}$ ' '') &  & candidature   & (km s$^{-1}$) \\
\noalign{\smallskip}
\hline
\noalign{\smallskip}
SIPS0007-2458  &  0  7  7.800 & -24 58  3.80 &  13.11$\pm$0.02 & IC     & 18 \\
2MASS0020-2346 &  0 20 23.155 & −23 46  5.38 &  12.35$\pm$0.02 & YD    & 12 \\
DENIS0021-4244 &  0 21  5.896 & -42 44 43.33 &  13.52$\pm$0.03 & IC, CA & 11 \\
SIPS0027-5401  &  0 27 23.240 & −54  1 46.20 &  12.36$\pm$0.02 & HY     & 27 \\
SIPS0153-5122  &  1 53 11.430 & −51 22 24.99 &  13.45$\pm$0.03 & IC, HY & 14 \\
SIPS0214-3237  &  2 14 45.440 & -32 37 58.20 &  14.01$\pm$0.02 & HY     & 18 \\
SIPS0235-0711  &  2 35 49.470 &  −7 11 21.90 &  12.45$\pm$0.03 & HY     & 22 \\
2MASS0334-2130 &  3 34 10.657 & −21 30 34.35 &  11.91$\pm$0.02 & IC     & $<$10$^{a}$ \\
SIPS2039-1126  & 20 39 13.081 & -11 26 52.30 &  13.79$\pm$0.03 & PL     & $>$15 \\
SIPS2045-6332  & 20 45  2.278 & -63 32  5.30 &  12.62$\pm$0.03 & PL, CA & $>$15 \\
LEHPM4908      & 22 36 42.656 & -69 34 59.30 &  12.68$\pm$0.02 & HY     & - \\
2MASS2254-3228 & 22 54 58.110 & −32 28 52.20 &  13.58$\pm$0.03 & PL, CA & 19 \\
LEHPM6542      & 23 57 54.822 & −19 55 1.89  &  13.31$\pm$0.02 & HY     & 16 \\
\noalign{\smallskip}
\hline
\noalign{\smallskip}
\end{tabular}
\end{center}
$^{1}$ MG to which targets are kinematic candidate (Paper II). HY= Hyades MG; 
SI= Ursa Major group; CA= Castor MG; PL= Pleiades; IC = IC 2391 MG; YD= Other young disk object.\\
$^{2}$ From paper II.\\
$^{a}$ Measused with low S/N.\\
\end{flushleft}
\end{table*}

\begin{table*}
\caption[]{B: Properties for sample B 
\label{tab:sampleb}}
\begin{flushleft}
\begin{center}
\small
\begin{tabular}{lllcccc}
\noalign{\smallskip}
\hline
\noalign{\smallskip}
Name & $\alpha$ (2000) & $\delta$ (2000) & J-mag & Moving group   \\
  & (h m s) & ($^{\rm o}$ ' '') & &  candidature   \\
\noalign{\smallskip}
\hline
\noalign{\smallskip}
2MASS0814-4020 & 08 14 35.46 & -40 20 49.26 & 14.356$\pm$0.023 & HY \\
2MASS1146-4754 & 11 46 51.04 & -47 54 38.17 & 14.897$\pm$0.042 & SI \\
2MASS1236-6536 & 12 36 32.38 & -65 36 35.6  & 15.277$\pm$0.057 & PL, CA, IC, SI \\
2MASS1326-5022 & 13 26 53.48 & -50 22 27.04 & 14.715$\pm$0.037 & IC, CA \\
2MASS1433-5148 & 14 33 41.95 & -51 48 03.70 & 14.206$\pm$0.034 & CA, PL, SI, IC \\
2MASS1557-4350 & 15 57 27.39 & -43 50 21.47 & 14.224$\pm$0.028 & PL, CA, IC \\
2MASS1618-3214 & 16 18 08.92 & -32 14 36.17 & 14.920$\pm$0.040 & IC, CA, SI \\
2MASS1734-1151 & 17 34 30.53 & -11 51 38.83 & 13.110$\pm$0.028 & PL \\
2MASS1736-0407 & 17 36 56.09 & - 4 07 25.84 & 15.516$\pm$0.070 & SI, CA \\
2MASS1745-1640 & 17 45 34.66 & -16 40 53.81 & 13.646$\pm$0.026 & SI, CA, HY \\
2MASS1756-4518 & 17 56 29.63 & -45 18 22.47 & 12.386$\pm$0.019 & CA \\
2MASS1909-1937 & 19 09 08.21 & -19 37 47.96 & 14.520$\pm$0.026 & CA \\
\noalign{\smallskip}
\hline
\noalign{\smallskip}
\end{tabular}
\end{center}
\end{flushleft}
\end{table*}

\begin{table*}
\caption[]{Details of observing runs 
\label{tab:obs}}
\begin{flushleft}
\begin{center}
\small
\begin{tabular}{ccccccc}
\noalign{\smallskip}
\hline
\noalign{\smallskip}
Number & Date & Telescope & Instrument & Spect. range & Dispersion    \\
  &  &           &              & (\AA)   & (\AA) \\
\noalign{\smallskip}
\hline
\noalign{\smallskip}
1 & 28/03-21/06/08 & ESO-VLT-U2 & UVES$^{1}$ & 6700-10425$^{2}$ & 0.027-0.041 \\
2 & 01-09-2009-09/01/2010 & ESO-VLT-U2 & UVES$^{1}$ & 5700-7530 \& 7650-9470  & 0.027-0.041\\
3 & 03-05/05/2010 & 6.5~m Baade-Maguellan & IMACS Short-Camera$^{3}$ & 6550-10000 & 1.98  \\
4 & 15-16/02/2011 & 6.5~m Baade-Maguellan & IMACS Short-Camera$^{3}$ & 4300-10800 & 1.97  \\
\noalign{\smallskip}
\hline
\noalign{\smallskip}
\end{tabular}
\end{center}
$^{1}$ UVES: Ultraviolet and Visual Echelle Spectrograph.\\
$^{2}$ Effective range.\\
$^{3}$ IMACS: The Inamori Magellan Areal Camera and Spectrograph.\\
\end{flushleft}
\end{table*}

\begin{table*}
\caption[]{Sample A: Model fit results for M3-M7.5 objects. Spectroscopic spectral types are from Paper II.
 The equivalent effective temperatures for the spectral types were taken from Reyl\'e et al. (2011).
 $\Delta$T$_{\rm eff}$= 100 K, $\Delta$$log~g$= 0.5 cm s$^{-2}$.  
\label{tab:resulta}}
\begin{flushleft}
\begin{center}
\small
\begin{tabular}{lcccccccccccccc}
\noalign{\smallskip}
\hline
\noalign{\smallskip}
 & \multicolumn{3}{c}{$SpT$} & \multicolumn{5}{c}{NextGen} & \multicolumn{4}{c}{Semi-empirical model} & \\
\noalign{\smallskip}
Object & $SpT_{Spec}$ & $T_{\rm eff}SpT$ & & $T_{\rm eff}$ & $log~g$ & [M/H] & $S_{min}$ & & $T_{\rm eff}$ & $log~g$ & [M/H] & $S_{min}$ & best fit$^{1}$ \\
       &       & (K)         & & (K) & (cm s$^{-2}$) & dex   &   & & (K) & cm s$^{-2}$ & dex &  & \\
\hline
\noalign{\smallskip}
GL876          & M4.0 & 3100 & & 3400 & 5.5 & 0.0 & 2.9 & & - & - & - & - & NextGen \\ 
SIPS0007-2458  & M7.5 & 2550 & & 2700 & 4.5 & -1.0 & 7.0  & & 2700 & 4.5 & 0.0 & 6.2  & s.-e. \\
2MASS0020-2346 & M6.0 & 2800 & & 2800 & 5.5 & -0.5 & 4.8  & & 2800 & 5.0 & 0.0 & 4.56 & s.-e. \\
SIPS0027-5401  & M6.5 & 2650 & & 2700 & 4.5 & -1.0 & 3.35 & & 2700 & 4.5 & 0.0 & 2.70 & s.-e. \\
SIPS0153-5122  & M6.0 & 2800 & & 2800 & 5.5 & -0.5 & 2.42 & & 2800 & 5.0 & 0.0 & 2.31 & s.-e. \\
SIPS0214-3237  & M6.5 & 2650 & & 2800 & 5.0 & -1.0 & 1.44 & & 2800 & 5.0 & 0.0 & 1.2  & s.-e. \\
SIPS0235-0711  & M6.0 & 2800 & & 2700 & 5.0 & -1.0 & 0.97 & & 2700 & 4.5 & 0.0 & 0.91 & s.-e. \\
2MAS0334-2130  & M4.5-M6.0 & 3000-2800 & & 2800 & 5.5 & -0.5 & 4.65 & & 2800 & 5.0 & 0.0 & 5.65 & $^{a}$    \\
LEHPM4908      & M6.0 & 2800 & & 2800 & 4.5 & -1.0 & 7.02 & & 2900 & 5.0 & 0.0 & 5.70 & s.-e. \\
2MASS2254-3228 & M5.5 & 2850 & & 2700 & 5.0 & -1.0 & 1.37 & & 2700 & 4.5 & 0.0 & 1.25 & s.-e. \\
LEHPM6542      & M6.0 & 2800 & & 2800 & 5.0 & -0.5 & 13.5 & & 2700 & 4.5 & 0.0 & 4.03 & s.-e. \\
\noalign{\smallskip}
\hline
\noalign{\smallskip}
\end{tabular}
\end{center}
$^{1}$ NextGen or semi-empirical (s.-e.) models.\\
$^{a}$ both models are equally likely. \\
\end{flushleft}
\end{table*}

\begin{table*}
\caption[]{Sample B:  Model fit results for M3-M7.5 objects. The equivalent effective temperature
 for the spectral types were taken from Reyl\'e et al. (2011).
  $\Delta$T$_{\rm eff}$= 100 K, $\Delta$$log~g$= 0.5 cm s$^{-2}$. 
\label{tab:resultb}}
\begin{flushleft}
\begin{center}
\scriptsize
\begin{tabular}{lccccccccccccccc}
\noalign{\smallskip}
\hline
\noalign{\smallskip}
 & \multicolumn{4}{c}{$SpT$} & \multicolumn{5}{c}{NextGen} & \multicolumn{4}{c}{Semi-empirical model} & \\
\noalign{\smallskip}
Object & $SpT_{Phot}$ & $SpT_{Spec}$ & $T_{\rm eff}Sp$ & & $T_{\rm eff}$ & $log~g$ & [M/H] & $S_{min}$ & & $T_{\rm eff}$ & $log~g$ & [M/H] & $S_{min}$ & best fit \\
       &       & (K)   & (K)      & & (K) & (cm s$^{-2}$) & dex   &  & & (K) & cm s$^{-2}$ & dex &  & \\
\hline
\noalign{\smallskip}
2MASS1146-4754 & M8.5 & M7-8 & $\sim$2500  & & 2800 & 5.5 &-0.5 & 0.39 & & 2700 & 4.5 & 0.0 & 0.18 & s.-e. \\
2MASS1236-6536 & M7.5 & M4.0 & 3100       & & 3200 & 5.5 & 0.0 & 0.64 & & 2900 & 4.0 & 0.0 & 1.14 & NextGen \\
2MASS1326-5022 & M9.0 & M7.5 & 2550       & & 2800 & 5.0 & 0.0 & 0.57 & & 2700 & 4.0 & 0.0 & 0.22 & s.-e. \\
2MASS1433-5148 & M9.0 & M6-7 & $\sim$2650 & & 3000 & 5.5 & 0.0 & 0.58 & & 2900 & 5.0 & 0.0 & 0.27 & s.-e. \\
2MASS1618-3214 & M7.0 & M6.5 & 2650       & & 2800 & 5.5 &-0.5 & 0.69 & & 2700 & 4.5 & 0.0 & 0.41 & s.-e. \\
2MASS1736-0407 & M8.5 & M4-5 & $\sim$3000 & & 3200 & 5.0 & 0.0 & 0.49 & & 2900 & 4.0 & 0.0 & 0.55 & NextGen  \\
\noalign{\smallskip}
\hline
\noalign{\smallskip}
\end{tabular}
\end{center}
\end{flushleft}
\end{table*}

\begin{table*}
\caption[]{Sample A: Model fit results for $\sim$M8 object. First line refers to fit results when use NextGen, DUSTY and COND models while second line refers to fit results when use the semi-empirical models based on NextGen, DUSTY and COND models respectively. 
 Spectroscopic spectral types are from Paper II.
  $\Delta$T$_{\rm eff}$= 100 K, $\Delta$$log~g$= 0.5 cm s$^{-2}$.
\label{tab:resultab}}
\begin{flushleft}
\begin{center}
\scriptsize
\begin{tabular}{lccccccccccccccc}
\noalign{\smallskip}

\hline
\noalign{\smallskip}
 & \multicolumn{2}{c}{$SpT$} & \multicolumn{4}{c}{NextGen} & \multicolumn{4}{c}{DUSTY} & \multicolumn{4}{c}{COND} & \\
Object & $SpT_{Spec}$ & $T_{\rm eff}SpT$ &  & $T_{\rm eff}$ & $log~g$ & $S_{min}$ & & $T_{\rm eff}$ & $log~g$ & $S_{min}$ & & $T_{\rm eff}$ & $log~g$ & $S_{min}$ & best fit \\   
 &    &          (K)     & & (K) & (cm s$^{-2}$) & & &  (K) & cm s$^{-2}$ & & & (K) & cm s$^{-2}$ & & \\

\hline
\noalign{\smallskip}
SIPS2039-1126 & M8.0 & ~2500 & & 2800 & 5.5 & 22.45 & & 2600 & 5.0 & 39.96 & & 2600 & 4.5 & 40.12 & \\
Semi-empirical & & & & 2700 & 4.5 & 22.90 & & 2600 & 4.5 & 20.00 & & 2600 & 4.5 & 20.01 & s.-e. DUSTY \\
\noalign{\smallskip}
\hline
\noalign{\smallskip}
\end{tabular}
\end{center}
\end{flushleft}
\end{table*}

\begin{table*}
\caption[]{Sample B: Model fit results for $\sim$M8 objects. 
 First line refers to fit results when use NextGen, DUSTY and COND models while second
 line refers to fit results when use the semi-empirical models based on NextGen, DUSTY and COND models respectively.
  $\Delta$T$_{\rm eff}$= 100 K, $\Delta$$log~g$= 0.5 cm s$^{-2}$.
\label{tab:resultbb}}
\begin{flushleft}
\begin{center}
\scriptsize
\begin{tabular}{lccccccccccccccc}
\noalign{\smallskip}
\hline
\noalign{\smallskip}
 & \multicolumn{2}{c}{$SpT$} & \multicolumn{4}{c}{NextGen} & \multicolumn{4}{c}{DUSTY} & \multicolumn{4}{c}{COND} & \\
Object & $SpT_{Phot}$ & $SpT_{Spec}$ & & $T_{\rm eff}$ & $log~g$ & $S_{min}$ & & $T_{\rm eff}$ & $log~g$ & $S_{min}$ & & $T_{\rm eff}$ & $log~g$ & $S_{min}$ & best fit \\
    &    &               & & (K) & (cm s$^{-2}$) & & & (K) & cm s$^{-2}$ & & & (K) & cm s$^{-2}$ & & \\

\hline
\noalign{\smallskip}
2MASS0814-4020 & L4.0 & M7-8 & & 2700 & 5.0 & 0.74 & & 2600 & 5.5 & 1.15 & & 2600 & 5.0 & 1.22 & \\
Semi-empirical & & & & 2700 & 4.5 & 0.35 & & 2600 & 4.5 & 0.28 & & 2600 & 4.5 & 0.28 & s.-e. DUSTY $^{1,2}$\\

2MASS1557-4350 & L0.0 & M7.5-8.5 & & 2800 & 4.5 & 0.74 & & 2600 & 5.5 & 0.93 & & 2400 & 5.5 & 1.11 \\
Semi-empirical & & & & 2800 & 4.0 & 0.31 & & 2500 & 4.0 & 0.29 & & 2600 & 4.0 & 0.29 & s.-e. DUSTY $^{1}$\\

2MASS1756-4518 & M7.5 & M8-9 & & 2600 & 5.5 & 0.21 & & 2600 & 5.5 & 0.18 & & 2600 & 5.0 & 0.20 &  \\
Semi-empirical & & & & 2600 & 4.0 & 0.09 & & 2600 & 5.5 & 0.08 & & 2600 & 5.0 & 0.10 & s.-e. DUSTY \\

\noalign{\smallskip}
\hline
\noalign{\smallskip}
\end{tabular}
\end{center}
$^{1}$ DUSTY and COND have the same $S_{min}$, but DUSTY model is more appropriate in the temperature range.\\
$^{2}$ We could appreciate parts of the spectrum were NextGen was the best fit.\\ 
\end{flushleft}
\end{table*}

\begin{table*}
\caption[]{Sample A: Model fit results for M8.5-M9.5 objects. 
 First line refers to fit results when use DUSTY and COND models while second
 line refers to fit results when use the semi-empirical models based on DUSTY and COND models respectively.
Spectroscopic spectral types are from Paper II.
  The equivalent effective temperature for the spectral types were taken from Dahn et al. (2002).
  $\Delta$T$_{\rm eff}$= 100 K, $\Delta$$log~g$= 0.5 cm s$^{-2}$.
\label{tab:resultac}}
\begin{flushleft}
\begin{center}
\small
\begin{tabular}{lcccccccccccccc}
\noalign{\smallskip}
\hline
\noalign{\smallskip}
 & \multicolumn{3}{c}{$SpT$} & \multicolumn{5}{c}{DUSTY} & \multicolumn{4}{c}{COND} & \\
\noalign{\smallskip}
Object & $SpT_{Spec}$ & $T_{\rm eff}SpT$ & & $T_{\rm eff}$ & $log~g$ & [M/H] & $S_{min}$ & & $T_{\rm eff}$ & $log~g$ & [M/H] & $S_{min}$ & best fit \\
       &       & (K)         & & (K) & (cm s$^{-2}$) & dex   &   & & (K) & cm s$^{-2}$ & dex &  & \\
\hline
\noalign{\smallskip}
DENIS0021-4244 & M9.5 & 2300-2400 & & 2000 & 5.5 & 0.0 & 7.85 & & 2400 & 5.5 & 0.0 & 13.40 & \\
Semi-empirical &  &  & & 2400 & 4.0 & 0.0 & 5.89 & & 2500 & 4.0 & 0.0 & 6.62 & s.-e. DUSTY \\
SIPS2045-6332  & M8.5 & 2300-2400 & & 2000 & 5.5 & 0.0 & 11.39 & & 2400 & 4.0 & 0.0 & 17.19 &  \\
Semi-empirical &  & & & 2400 & 4.0 & 0.0 & 8.14 & & 2500 & 4.0 & 0.0 & 10.30 & s.-e. DUSTY \\
LP944-20       & M9.0 & 2300-2400 & & 2000 & 5.5 & 0.0 & 0.42 & & 2400 & 5.5 & 0.0 & 0.57 &  \\
Semi-empirical & & & & 2400 & 4.0 & 0.0 & 0.29 & & 2400 & 4.0 & 0.0 & 0.35 & s.-e. DUSTY \\

\noalign{\smallskip}
\hline
\noalign{\smallskip}
\end{tabular}
\end{center}
\end{flushleft}
\end{table*}

\begin{table*}
\caption[]{Sample B: Model fit results for $>$M8.5 objects. 
 First line refers to fit results when use DUSTY and COND models while second
 line refers to fit results when use the semi-empirical models based on DUSTY and COND models respectively.
The equivalent effective temperature
 for the spectral types were taken from Reyl\'e et al. (2011).
  $\Delta$T$_{\rm eff}$= 100 K, $\Delta$$log~g$= 0.5 cm s$^{-2}$.
\label{tab:resultbc}}
\begin{flushleft}
\begin{center}
\tiny
\begin{tabular}{lccccccccccccccc}
\noalign{\smallskip}
\hline
\noalign{\smallskip}
 & \multicolumn{3}{c}{$SpT$} & \multicolumn{5}{c}{DUSTY} & \multicolumn{4}{c}{COND} & \\
Object & $SpT_{Phot}$ & $SpT_{Spec}$ & $T_{\rm eff}SpT$ & & $T_{\rm eff}$ & $log~g$ & [M/H] & $S_{min}$ & & $T_{\rm eff}$ & $log~g$ & [M/H] & $S_{min}$ & best fit \\
    &    &       & (K)         & & (K) & (cm s$^{-2}$) & dex   &   & & (K) & cm s$^{-2}$ & dex &  & \\
\hline
\noalign{\smallskip}
2MASS1734-1151 & L0.0  & M9.0  & 2300-2500 & & 2600 & 5.5 & 0.0 & 1.34 &  & 2600 & 5.0 & 0.0 & 1.43 &  \\
Semi-empirical & & & & & 2400 & 4.0 & 0.0 & 0.79 & & 2400 & 4.0 & 0.0 & 0.86 & s.-e. DUSTY \\   
2MASS1745-1640 & L8.0  & M9-L2 & $\sim$2300-2000 & & 2000 & 5.5 & 0.0 & 0.06 & & 2200 & 5.5 & 0.0 & 0.07 &  \\
Semi-empirical & & & & & 2000 & 5.5 & 0.0 & 0.05 & & 2200 & 5.5 & 0.0 & 0.05 & s.-e. DUSTY \\
2MASS1909-1937 & M9.0  & L0.0  & $\sim$2200 & & 2100 & 3.5 & 0.0 & 0.55 & & - & - & - & - &  \\
Semi-empirical & & & & & 2000/2100 & 6.0/3.5 & 0.0 & 0.45 & & - & - & - & - & s.-e. DUSTY ? \\
\noalign{\smallskip}
\hline
\noalign{\smallskip}
\end{tabular}
\end{center}
\end{flushleft}
\end{table*}

\begin{table*}
 \centering
\caption[]{MG membership parameters. For each target, we give the result 
 obtained for each criterion used.
  N/A= non applicable criterion in this case; Y= the age parameter agrees with the MG membership;
 N= the age parameter does not agree with the MG membership; when more than one membership was
  possible in column 5, the MG name that the criteria determined as possible is given. Where 
 HY is Hyades, SI is Ursa Major, IC is IC 2391, CA is Castor and PL is Pleiades MG and YD is other young disk
 object. When interrogation appears after MG name
 we indicate that the membership probability is fewer than for other MGs or in the case of 2MASS1734-1151 and 2MASS1909-1937,
 that the criteria are not conclusive.\\
\label{tab:resultfinal}}
\begin{flushleft}
\begin{center}
\scriptsize
\begin{tabular}{lcccccccccc}
\noalign{\smallskip}
\hline
\noalign{\smallskip}
Object & $EW$(Na~{\sc i})$^{1}$ & $EW$(H$\alpha$)$^{2}$ & age$^{2}$ & MG memb. & M. from & M. from  & M. from & M. from & M. from & Final MG \\
       & (\AA) & (\AA) & (Myr) & (Kinem/Astrom)$^{3}$ & $EW$(Na~{\sc i})$^{1}$ & $EW$(H$\alpha$)$^{2}$ & $log~g$$^{4}$ & Li~{\sc i}$^{5}$ & $v\sin{i}$$^{6}$ & \\
\hline
\noalign{\smallskip}
SIPS0007-2458  & 8.2 & 11.0   & 1-300     & IC             & N  & Y  & Y  & N  & Y & {\it YD} \\
2MASS0020-2346 & 8.1 &  6.1   & $\sim$300 & YD            & Y  & Y  & Y  & ?  & Y & {\it YD} \\
DENIS0021-4244 & 8.4 & $^{a}$ & N/A        & IC, CA         & CA & N/A & Y  & -  & N & {\it CA} \\
SIPS0027-5401  & 7.7 &  8.1   & $\sim$300 & HY             & N/A & Y  & Y  & Y  & Y & {\it HY} \\
SIPS0153-5122  & 8.0 &  7.5   & $\sim$300 & IC, HY         & HY & HY & HY & HY & Y & {\it HY} \\
SIPS0214-3237  & 7.9 &  8.4   & $\sim$300 & HY             & N/A & Y  & Y  & Y  & Y & {\it HY}\\
SIPS0235-0711  & 7.9 &  7.7   & $\geq$300 & HY             & N/A & Y  & Y  & Y  & Y & {\it HY} \\
2MASS0334-2130 & 7.3 & 11.5   & 1-300     & IC             & N  & Y  & N  & N  & ? & {\it YD }\\
SIPS2045-6332  & 6.3 &  2.1   & N/A        & PL, CA         & CA & N/A & Y  & Y  & ? & {\it CA }\\
SIPS2039-1126  & 7.3 &   -    & -         & PL             & N  & -  & Y  & Y  & Y & {\it PL}\\
LEHPM4908      & 7.3 &  6.5   & $\sim$300 & HY             & N/A & Y  & Y  & Y  & - & {\it HY }\\
2MASS2254-3228 & 8.5 &  5.8   & $\sim$300 & PL, CA         & CA & CA & Y  & -  & Y & {\it CA }\\
LEHPM6542      & 7.6 &  4.9   & $\geq$300 & HY             & N/A & Y  & Y  & Y  & Y & {\it HY }\\
               & & &                &    &    & & & & \\
2MASS0814-4020 & 9.0 &  7.2    & $\geq$300 & HY              & N/A     & Y      & Y      & N/A & N/A & {\it HY}\\
2MASS1146-4754 & 7.9 &  5.2    & $\geq$300 & SI              & N/A     & Y      & Y      & N/A & N/A & {\it SI}\\
2MASS1236-6536 & 5.8 &  1.0    & 300-1200  & PL, IC, CA, SI  & CA, SI & CA, SI & N      & N/A & N/A & {\it CA, SI} \\
2MASS1326-5022 & 4.2 & 23.6    & 1-100     & IC, CA          & IC     & IC     & Y      & N/A & N/A & {\it IC}\\
2MASS1433-5148 & 7.9 & 19.8    & 1-100     & PL, IC, CA, SI  & CA, SI & PL, IC & SI     & N/A & N/A & {\it CA?, SI} \\
2MASS1557-4350 & 4.0 & 16.4    & 1-100     & PL, IC, CA      & PL, IC & N/A     & Y      & N/A & N/A & {\it PL, IC} \\
2MASS1618-3214 & 8.0 &  7.8    & 1-300     & IC, CA, SI      & CA, SI & Y      & Y      & N/A & N/A & {\it IC?, CA, SI}\\
2MASS1734-1151 & 7.7 &  5.5    & N/A        & PL              & N      & N/A     & Y      & N/A & N/A & {\it PL?}\\
2MASS1736-0407 & 6.1$^{c}$ &  4.2    & 50-300    & CA, SI          & N/A     & Y      & Y      & N/A & N/A & {\it CA, SI}\\
2MASS1745-1640 & 5.6 &  1.4    & $\geq$300  & CA, SI, HY      & Y      & N/A     & HY     & N/A & N/A & {\it CA?, SI?, HY} \\
2MASS1756-4518 & 7.5 &  3.2    & N/A        & CA              & N/A     & N/A     & Y      & N/A & N/A & {\it CA}\\
2MASS1909-1937 & 6.3 & $^{a}$  & N/A        & CA              & N/A     & N/A     & -      & N/A & N/A & {\it CA?}\\
\noalign{\smallskip}
\hline
\noalign{\smallskip}
\end{tabular}
\end{center}
$^{1}$ Equivalenth width of the Na~{\sc i} doublet as expained in Section 5.1.\\
$^{2}$ See Section 5.2.\\
$^{3}$ MG membership from kinematic (sample A) or from photometric and astrometric criteria (sample B). See Section 2.\\
$^{4}$ From value obtained in the synthetic fits, given in columns 5 and 9 of Tables~\ref{tab:resulta}, ~\ref{tab:resultb} and ~\ref{tab:resultab}.\\
$^{5}$ Only applicable for high resolution data. See Section 5.3.\\
$^{6}$ From Paper II.\\
$^{a}$ Absorption line filled in with emission.\\
$^{b}$ Absorption line.\\
$^{c}$ Measured with a cosmic ray in the line.\\

\end{flushleft}
\end{table*}

\section*{Acknowledgments}
 M.C. G\'alvez-Ortiz acknowledges financial support by the Spanish MICINN under 
 the Consolider-Ingenio 2010 Program grant CSD2006-00070: First Science with 
 the GTC (http://www.iac.es/consolider-ingenio-gtc). She also acknowledges the 
 support of a JAE-Doc CSIC fellowship co-funded with the European Social Fund
 under the program {\em ”Junta para la Ampliaci\'on de Estudios”}.
 M.K. Kuznetsov and Ya.V. Pavlenko work was supported by EU PF7 Marie Curie 
 Initial Training Networks (ITN) ROPACS project (GA N 213646).
 Financial support was also provided by the Spanish Ministerio de Ciencia e Innovaci\'on 
 and Ministerio de Econom\'ia y Competitividad under AyA2011- 30147-C03-03 grant.
 S.L.F. acknowledges funding support from the ESO-Government of Chile Mixed
 Committee 2009, and from the GEMINI--CONICYT grant \# 32090014/2009.
 J.S.J. acknowledges funding by Fondecyt through grant 3110004 and partial support 
 from CATA (PB06, Conicyt), the GEMINI-CONICYT FUND and from the Comit\'e Mixto ESO-GOBIERNO DE CHILE.
 Financial support was also provided by the Mexican research council (CONACYT) grant.

 Based on observations made with ESO Telescopes, with UVES high-resolution optical spectrograph at 
 VLT Kueyen 8.2-m telescope at Paranal Observatory (Chile) in programmes 081.C-0222(A) and 084.C-0403(A).
 And Observations obtained at the Magellan telescope at Las Campanas (Chile) Observatory.

\bsp

\label{lastpage}

\appendix
\section{Tables}
\begin{table*}
\caption[] {Compilation of all VLM moving group candidates: the fith column describes the method or methods 
 from which membership has been derived. When the word probable is used we mean that
 kinematics is not supported by $v\sin{i}$ criterion at paper II.
 Spectral types given are from spectral indices (calculated in Paper II and here) except when marked.
\\
\label{tab:comptod}}
\begin{flushleft}
\begin{center}
\small
\begin{tabular}{lllll}
\noalign{\smallskip}
\hline
\noalign{\smallskip}
Name & $\alpha$ (2000) & $\delta$ (2000) & $SpT$ & Note \\
  & (h m s) & ($^{\rm o}$ ' '') &  &  \\
\noalign{\smallskip}
\hline
\noalign{\smallskip}
Hyades         &              &              &      &  \\
SIPS0004-5721  &  0  4 18.970 & -57 21 23.30 & M7.0 & also IC candidate; kinematic\\
SIPS0027-5401  &  0 27 23.240 & -54  1 46.20 & M6.0 & kinematics + age features\\
2MASS0123-3610 &  1 23  0.506 & -36 10 30.67 & M4.5 & kinematics \\
SIPS0153-5122  &  1 53 11.430 & -51 22 24.99 & M6.0 & kinematics + age features\\
SIPS0214-3237  &  2 14 45.440 & -32 37 58.20 & M6.5 & kinematics + age features\\
SIPS0235-0711  &  2 35 49.470 &  -7 11 21.90 & M6.0 & kinematics + age features\\
SIPS0440-0530  &  4 40 23.328 &  -5 30  7.85 & M7.5 & kinematics \\
2MASS0600-3314 &  6  0 33.750 & -33 14 26.84 & M7.0$^{phot}$ & probable member from kinematics \\
2MASS0814-4020 &  8 14 35.46  & -40 20 49.26 & M7-8 & photometry + astrometry + age features\\
2MASS1745-1640 & 17 45 34.66  & -16 40 53.81 & M9-L2 & also CA and SI candidate; photometry + astrometry + age features\\
SIPS2014-2016  & 20 14  3.523 & -20 16 21.30 & M7.5 & kinematics \\
2MASS2031-5041 & 20 31 27.495 & -50 41 13.49 & M5.0 & kinematics \\
SIPS2049-1716  & 20 49 52.610 & -17 16  7.80 & M6.5 &  kinematics \\
SIPS2100-6255  & 21  0 30.227 & -62 55  7.31 & M5.0$^{phot}$ & kinematics \\
DENIS2200-3038 & 22  0  2.022 & -30 38 32.71 & M9.0 & probable member from kinematics \\
SIPS2200-2756  & 22  0 16.838 & -27 56 29.70 & M6.0 & probable member from kinematics \\
2MASS2207-6917 & 22  7 10.313 & -69 17 14.25 & M6.5 & kinematics \\
2MASS2231-4443 & 22 31  8.657 & -44 43 18.43 & M4.5 & kinematics \\
LEHPM4908      & 22 36 42.656 & -69 34 59.30 & M5.5 & kinematics + age features\\
2MASS2311-5256 & 23 11 30.330 & -52 56 30.17 & M5.5 & kinematics \\
SIPS2318-4919  & 23 18 45.952 & -49 19 17.79 & M6.5 & kinematics \\
SIPS2322-6357  & 23 22  5.332 & -63 57 57.60 & M6.5 & kinematics \\
SIPS2347-1821  & 23 47 16.662 & -18 21 50.60 & M6.5 & kinematics \\
SIPS2350-6915  & 23 50  3.948 & -69 15 24.39 & M6.5 & kinematics \\
LEHPM6375      & 23 52 49.138 & -22 49 29.54 & M6.5 & probable member from kinematics \\
LEHPM6542      & 23 57 54.822 & -19 55  1.89 & M6.0 & kinematics + age features\\
Ursa Major     &              &              &      &  \\
2MASS1146-4754 & 11 46 51.04  & -47 54 38.17 & M7-8 & photometry + astrometry + age features\\
2MASS1236-6536 & 12 36 32.38  & -65 36 35.6  & M4.0 & also CA candidate; photometry + astrometry + age features\\
2MASS1433-5148 & 14 33 41.95  & -51 48 03.70 & M6-7 & also CA candidate; photometry + astrometry + age features\\ 
2MASS1618-3214 & 16 18 08.92  & -32 14 36.17 & M6.5 & also CA candidate; photometry + astrometry + age features\\
2MASS1736-0407 & 17 36 56.09  & - 4 07 25.84 & M4-5 & also CA candidate; photometry + astrometry + age features\\
2MASS1745-1640 & 17 45 34.66  & -16 40 53.81 & M9-L2 & also HY and CA candidate; photometry + astrometry + age features\\
Castor         &              &              &      &  \\
DENIS0021-4244 &  0 21  5.896 & -42 44 43.33 & M9.5 & kinematics + age features\\
2MASS1236-6536 & 12 36 32.38  & -65 36 35.6  & M4.0 & also SI candidate; photometry + astrometry + age features\\
2MASS1433-5148 & 14 33 41.95  & -51 48 03.70 & M6-7 & also SI candidate; photometry + astrometry + age features\\
2MASS1618-3214 & 16 18 08.92  & -32 14 36.17 & M6.5 & also SI candidate; photometry + astrometry + age features\\
2MASS1736-0407 & 17 36 56.09  & - 4 07 25.84 & M4-5 & also SI candidate; photometry + astrometry + age features\\ 
2MASS1745-1640 & 17 45 34.66  & -16 40 53.81 & M9-L2 & also HY and SI candidate; photometry + astrometry + age features\\
2MASS1756-4518 & 17 56 29.63  & -45 18 22.47 & M8-9 & photomerty + astrometry + age features\\
2MASS1909-1937 & 19 09 08.21  & -19 37 47.96 & L0.0 & photometry + astrometry \\
SIPS2000-7523  & 20  0 48.171 & -75 23  6.58 & M8.0 & probable member from kinematics\\
SIPS2045-6332  & 20 45  2.278 & -63 32  5.30 & M9.0 & kinematics + age features\\
SIPS2114-4339  & 21 14 40.928 & -43 39 51.20 & M6.5 & kinematics \\
2MASS2242-2659 & 22 42 41.294 & -26 59 27.23 & M5.5 & also HY candidate; kinematics \\
2MASS2254-3228 & 22 54 58.110 & -32 28 52.20 & M5.5 & kinematics + age features\\
Pleiades       &              &              &      &  \\
2MASS1557-4350 & 15 57 27.39  & -43 50 21.47 & M7.5-8.5 & also IC candidate; photometry + astrometry + age features\\
2MASS1734-1151 & 17 34 30.53  & -11 51 38.83 & M9.0 & photometry + astrometry\\
SIPS2039-1126  & 20 39 13.081 & -11 26 52.30 & M7.0 & kinematics + age features\\
HB2124-4228    & 21 27 26.133 & -42 15 18.39 & M7.5 & kinematics \\
IC 2391        &              &              &      &  \\
SIPS0004-5721  &  0  4 18.970 & -57 21 23.30 & M7.0 & also HY candidate; kinematics \\
2MASS1326-5022 & 13 26 53.48  & -50 22 27.04 & M7.5 & photometry + astrometry + age features\\
2MASS1557-4350 & 15 57 27.39  & -43 50 21.47 & M7.5-8.5 & also PL candidate; photometry + astrometry + age features\\
2MASS1618-3214 & 16 18 08.92  & -32 14 36.17 & M6.5 & also CA and SI candidate; photometry + astrometry + age features\\
SIPS2341-3550  & 23 41 47.497 & -35 50 14.40 & M7.0 & kinematics \\
\noalign{\smallskip}
\hline
\noalign{\smallskip}
\end{tabular}
\end{center}
\end{flushleft}
\end{table*}

\begin{table*}
\caption[] {Other probably young disk and old disk candidates. Notes as in previous table.
\label{tab:compothers}}
\begin{flushleft}
\begin{center}
\small
\begin{tabular}{lllll}
\noalign{\smallskip}
\hline
\noalign{\smallskip}
Name & $\alpha$ (2000) & $\delta$ (2000) & $SpT$ & Note \\
  & (h m s) & ($^{\rm o}$ ' '') &  &  \\
\noalign{\smallskip}
\hline
\noalign{\smallskip}
Other Young Disk &              &              &      &  \\
SIPS0007-2458  &  0  7  7.800 & -24 58  3.80 & M7.5 & kinematics + age features\\
2MASS0020-2346 &  0 20 23.155 & -23 46  5.38 & M6.0 & kinematics + age features\\
SIPS0039-2256  &  0 39 23.250 & -22 56 44.90 & M7.5 & probable YD from kinematics \\
DENIS0041-5621 &  0 41 35.390 & -56 21 12.77 & M7.5 & kinematics\\
SIPS0054-4142  &  0 54 35.300 & -41 42  6.20 & M5.0 & probable YD from kinematics\\
SIPS0109-0343  &  1  9 51.040 &  -3 43 26.30 & M9.0 & not conclusive from kinematics\\
LEHPM1289      &  1  9 59.579 & -24 16 47.82 & M6.0 & not conclusive from kinematics\\
SIPS0115-2715  &  1 15 26.610 & -27 15 54.10 & M5.0 & not conclusive from kinematics\\
SIPS0126-1946  &  1 26 49.980 & -19 46  5.90 & M6.0 & not conclusive from kinematics\\
LEHPM1563      &  1 27 31.956 & -31 40  3.18 & M8.5 & kinematics\\
SIPS0212-6049  &  2 12 33.580 & -60 49 18.40 & M6.5 & not conclusive from kinematics\\
2MASS0334-2130 &  3 34 10.657 & -21 30 34.35 & M4.5 & kinematics + age features\\
2MASS0429-3123 &  4 29 18.426 & -31 23 56.81 & M7.5 & not conclusive from kinematics\\
2MASS0502-3227 &  5  2 38.677 & -32 27 50.07 & M5.5 & not conclusive from kinematics\\
2MASS0528-5919 &  5 28  5.623 & -59 19 47.17 & M5.5 & not conclusive from kinematics\\
SIPS1039-4110  & 10 39 18.340 & -41 10 32.00 & M6.5$^{phot}$ & probable YD from kinematics\\
SIPS1124-2019  & 11 24 22.229 & -20 19  1.50 & M7.0$^{phot}$ & probable YD from kinematics\\
SIPS2049-1944  & 20 49 19.673 & -19 44 31.30 & M7.0 & kinematics \\
SIPS2128-3254  & 21 28 17.402 & -32 54  3.90 & M6.5 & kinematics \\
SIPS2321-6106  & 23 21 43.418 & -61  6 35.37 & M5.0 & kinematics \\
SIPS2343-2947  & 23 43 34.731 & -29 47  9.50 & M8.0 & kinematics \\
Old Disk       &              &              &      &  \\
2MASS0204-3945 &  2  4 18.036 & -39 45  6.48 & M7.0 & not conclusive from kinematics\\
2MASS0445-5321 &  4 45 43.368 & -53 21 34.56 & M7.5$^{phot}$ & kinematics\\
DENIS1250-2121 & 12 50 52.654 & -21 21 13.67 & M7.5 & kinematics\\
SIPS1329-4147  & 13 29  0.872 & -41 47 11.90 & M9.5$^{phot}$ & kinematics\\
SIPS1341-3052  & 13 41 11.561 & -30 52 49.60 & L0$^{phot}$ & kinematics\\
2MASS1507-2000 & 15  7 27.799 & -20  0 43.18 & M7.5 & kinematics\\
SIPS1632-0631  & 16 32 58.799 &  -6 31 45.30 & M8.5 & kinematics\\
SIPS1758-6811  & 17 58 59.663 & -68 11 10.50 & M5.0 & probable OD from kinematics\\
SIPS1949-7136  & 19 49 45.527 & -71 36 50.89 & M7.0 & kinematics\\
2MASS2001-5949 & 20  1 24.639 & -59 49  0.09 & M6.0 & probable OD from kinematics\\
2MASS2106-4044 & 21  6 20.896 & -40 44 51.91 & M6.0 & probable OD from kinematics \\
SIPS2119-0740  & 21 19 17.571 &  -7 40 52.50 & M7.0 & probable OD from kinematics \\
LEHPM4480      & 22 15 10.151 & -67 38 49.07 & M5.5 & probable OD from kinematics \\
2MASS2222-4919 & 22 22  3.684 & -49 19 23.45 & M6.5 & kinematics \\
\noalign{\smallskip}
\hline
\noalign{\smallskip}
\end{tabular}
\end{center}
\end{flushleft}
\end{table*}

\end{document}